\documentclass[fleqn,usenatbib]{mnras}

\usepackage{newtxtext,newtxmath}

\usepackage[T1]{fontenc}

\DeclareRobustCommand{\VAN}[3]{#2}
\let\VANthebibliography\thebibliography
\def\thebibliography{\DeclareRobustCommand{\VAN}[3]{##3}\VANthebibliography}

\usepackage{graphicx}	
\usepackage{amsmath}	
\usepackage{cleveref}
\usepackage{subcaption}
\usepackage{multirow}
\usepackage{float}
\usepackage{soul}

\crefformat{figure}{Fig.~#2#1#3}
\crefformat{equation}{Eq.~(#2#1#3)}
\crefformat{section}{section~#2#1#3}
\crefformat{table}{Tab.~#2#1#3}
\crefformat{appendix}{appendix~#2#1#3}
\crefmultiformat{figure}{Figs.~#2#1#3}{~and~#2#1#3}%
    {,~#2#1#3}{,~#2#1#3}
\crefrangeformat{figure}{Figs.~(#3#1#4--#5#2#6)}
\crefrangeformat{equation}{Eqs.~(#3#1#4--#5#2#6)}

\setstcolor{red}

\title[Star formation on the main sequence]{Star-formation variability on the star-forming main sequence from Cosmic Noon to the Epoch of Reionization}
\author[Bevins, Tacchella \& Simmonds]{
H. T. J. Bevins,$^{1,2,3,4}$\thanks{E-mail: h.bevins@imperial.ac.uk}
S. Tacchella,$^{1,2}$
C. Simmonds$^{1,2,5}$
\\
$^{1}$Cavendish Laboratory, University of Cambridge, Cambridge, CB3 0HE, UK\\
$^{2}$Kavli Institute for Cosmology, University of Cambridge, Cambridge, CB3 0HA, UK\\
$^{3}$Department of Physics, Imperial College London, Blackett Laboratory, Prince Consort Road, London, SW7 2AZ, UK\\
$^{4}$I-X Centre for AI in Science, Translation and Innovation Hub (I-HUB), Imperial College London, White City Campus, 84 Wood Lane, London, W12 0BZ, UK\\
$^{5}$Departamento de Astronomía, Universidad de Chile, Camino El Observatorio 1515, Las Condes, Santiago, Chile
}

\date{Accepted XXX. Received YYY; in original form ZZZ}

\pubyear{\the\year{}}

\begin{document}
\label{firstpage}
\pagerange{\pageref{firstpage}--\pageref{lastpage}}
\maketitle

\begin{abstract}
Star formation in galaxies is intrinsically stochastic, driven by physical processes operating across a wide range of scales. The scatter in the star-forming main sequence relation provides a window into this variability, but interpreting it in terms of underlying physical mechanisms remains challenging. We present a study of star-formation variability from cosmic noon to reionization ($z=3-8$) using power spectral density (PSD) models to characterize fluctuations in star formation rates. We use estimates of the intrinsic main-sequence scatter at six averaging timescales ($10-100$ Myr) from a catalogue of $\approx 17,000$ galaxies to constrain two PSD models, the Simple Harmonic Oscillator (\texttt{SHO}) and the Extended Regulator (\texttt{ExtReg}). We find that the regulator component of \texttt{ExtReg} is poorly constrained by present data. However, both the dynamical component of \texttt{ExtReg} and the single-component \texttt{SHO} model favour characteristic variability timescales of $\sim 10-30$ Myr, comparable to expected galactic dynamical and stellar feedback timescales. In the \texttt{SHO} model, most clearly at $z=3-4$, the inferred PSD power on $10$ Myr timescales decreases with stellar mass, indicating more bursty star formation in lower-mass galaxies. We find weak evidence for a transition from a two-component \texttt{ExtReg}-like PSD at lower redshift to a single-component \texttt{SHO}-like PSD at higher redshift in the lowest mass bin ($\log M_*/M_\odot = 8-8.5$), although the Bayes factors are small and high-redshift selection effects prevent strong conclusions. Overall, our results suggest that the observed $10-100$ Myr scatter of the high-redshift star-forming main sequence is governed primarily by short-timescale variability, consistent with galactic dynamical timescales.
\end{abstract}




\section{Introduction}
\label{sec:introduction}

Galaxy scaling relations, such as the correlation between star-formation rate (SFR) and stellar mass (the star-forming main sequence, SFMS), the stellar mass-metallicity relation, and the size-mass relation, encode how galaxies assemble their baryons through the processes of gas accretion, star formation, and feedback from stars and black holes within a cosmological framework. Thanks to the sensitivity and spectral leverage of \textit{JWST}, these relations can now be robustly traced to redshift $z\approx9$, merely $\simeq 600$ Myr after the Big Bang \citep[e.g.,][]{curti2024, Simmonds2025burstingattheseams, Scholtz2025, Allen2025, danhaive2025dawnofdiscs}. While establishing the relations themselves is increasingly feasible, the harder problem -- and the focus of this work -- is to infer how galaxies evolve through them: to connect the observed normalization, slope, and scatter of the SFMS to the underlying, time-variable physics that regulates star formation in galaxies. 

The SFMS, defined as the linear relation in logarithmic space between stellar mass (M$_*$) and SFR (or sSFR$\equiv$SFR/M$_*$), has been widely studied up to $z\sim 9$ \citep[e.g.,][]{Brinchmann2004,Daddi2007,Speagle2014,Popesso2023,Clarke2024,Simmonds2025burstingattheseams}. One of the key insights gained by studying the shape and normalization of the SFMS is understanding which mechanisms drive galaxy mass assembly across cosmic times: simple galaxy formation models predict that this relation follows accretion onto dark matter halos \citep{Bouche2010,Lilly2013,Rodriguez-Puebla2016,Tacchella2018}, a prediction supported by some observational studies \citep[e.g.,][]{Sandles2022,Simmonds2025burstingattheseams}. However, a secondary avenue to study is the scatter of the SFMS, which describes how galaxies are distributed around this relation. Importantly, the amount of scatter at a given averaging timescale encodes important information on how star formation takes place on global scales, and how stochastic this process is \citep{Gladders2013,Tacchella2016,Caplar2019,Wang2020,Iyer2020,McClymont2025,Wan2025}. 

Several studies focused on galaxies with M$_*$ above $10^9$ M$_{\odot}$ have found that the intrinsic scatter ($\sigma_{\rm MS}$) of the SFMS has only a weak dependence on stellar mass and redshift \citep[with $\sigma_{\rm MS}\leq 0.5$ dex;][]{Elbaz2007,Noeske2007,Behroozi2013,Schreiber2015,Kurczynski2016,Sandles2022}. However, interestingly, the scatter has been shown to increase at lower stellar masses (M$_*\lesssim 10^9$ M$_{\odot}$) in both observations \citep{Kauffmann2014,Weisz2014,Santini2017,Boogaard2018,Atek2022,Asada2024,Cole2025} and numerical simulations \citep{Shen2014,Dominguez2015,Iyer2020,Dome2024,McClymont2025}. This is the stellar mass regime where bursty star formation is expected to take place \citep{Weisz2012,Guo2015} due to the increased importance of stellar feedback and the sampling of individual star-forming regions \citep{Dekel1986,Stinson2007,Tacchella2020ExtReg,Dome2024}. Moreover, at high-redshift, burstiness has been explained by stochastic intergalactic medium (IGM) inflow  \citep{McClymont2025}, and shorter dynamical timescales \citep{Faucher-Giguere2018,Tacchella2020ExtReg}.

In mathematical terms, a galaxy’s star-formation history (SFH) can be described as ``bursty'' when it exhibits significant variability on short timescales relative to its overall dynamical timescale --- the characteristic timescale for gravity to drive coherent motion within the galaxy. A convenient way to characterise this variability is through the temporal power spectral density (PSD), which encodes the distribution of fluctuations in the SFH across timescales \citep{Caplar2019, Tacchella2020ExtReg, Munoz2026psd}. The PSD can be used to link different modes of variability to distinct physical processes: fluctuations on very short timescales ($\lesssim10$ Myr) reflect internal mechanisms such as giant molecular cloud formation, star formation, and stellar feedback \citep{Leitherer1999,Tan2000,Tasker2011,Faucher-Giguere2018,Benincasa2020}, while variations on intermediate ($\sim$10–100 Myr) and longer ($\gtrsim$100 Myr) timescales are dominated by large-scale feedback, gas cycling, and external drivers such as mergers \citep{Robertson2006,Semenov2017,Shin2023,Puskas2025, Kelson2020Gravity}. In particular, breaks in the PSD reveal the characteristic correlation timescales of the underlying processes, providing insight into how long different modes of star formation remain coherent.

In theoretical models it is usually possible to track the SFHs of individual galaxies across cosmic time, but this is not feasible observationally. Instead, SFHs must be inferred indirectly by decoding the integrated light of stellar populations. While the stellar continuum contains information about stars of different ages, additional effects from metallicity, dust attenuation, and nebular emission introduce significant degeneracies \citep{Papovich2001StellarPopulations, Walcher2011SED, Conroy2013PanchromaticSED, Schaerer2013lymanbreakgalaxies}. Even with high signal-to-noise, multi-wavelength spectroscopic data, sensitivity to stellar ages declines roughly logarithmically with look-back time, and young stars tend to outshine older populations \citep{Papovich2001StellarPopulations, Conroy2013PanchromaticSED, Whitaker2017, Williams2024, Sun2024, Papovich2023, Herard-Demanche2025, Reddy2026}. As a result, short-term fluctuations in the SFR cannot be recovered for individual galaxies at large look-back times. To constrain star-formation variability observationally, one must therefore turn to ensembles of galaxies, under the assumption that they share similar variability properties \citep{Weisz2012, Guo2015, Caplar2019}. It is thus essential to construct such ensembles carefully, for example by selecting galaxies at comparable cosmic time and stellar mass. In this context, the scatter of the star-forming main sequence provides a powerful statistical probe of star-formation variability across timescales.

In this work, we translate the observed, timescale-dependent scatter of the SFMS into constraints on the variability of star formation. We forward-model the SFMS at $z\simeq3$–8 using the observational constraints of \citet{Simmonds2025burstingattheseams}, who measured the intrinsic scatter from SFRs averaged over $t_{\rm avg}=10$–100\,Myr in stellar-mass and redshift bins. \citet{Simmonds2025burstingattheseams} found that this scatter is enhanced on shorter averaging timescales, providing evidence for bursty star formation at high redshift, consistent with recent independent results from \cite{Mitsuhashi2026}. Here, we interpret these measurements with a PSD framework, comparing a phenomenological Simple Harmonic Oscillator (\texttt{SHO}) model \citep{ForemanMackey2017SHO} to the physically motivated extended regulator (\texttt{ExtReg}) model \citep{Tacchella2020ExtReg} that couples inflow–equilibrium cycling of gas with a stochastic prescription for star formation driven by dynamical processes like the formation and disruption of giant molecular clouds. For each model we construct Gaussian-process SFHs \citep{Iyer2024ExtReg}, propagate them through the same averaging as the data, and emulate $\sigma_{\rm MS}(t_{\rm avg})$ with neural networks to enable likelihood-based inference with nested sampling. We validate the pipeline on mock observations, infer PSD parameters (amplitudes and correlation timescales), and compare Bayesian evidences across stellar mass and redshift, thereby identifying the processes and timescales that dominate SF variability during cosmic noon and the Epoch of Reionization.

This paper is structured as follows. In \Cref{sec:observations} we describe the observational dataset and the derivation of galaxy properties. In \Cref{sec:simulations} we introduce the framework for modelling stochastic star formation via PSDs and outline the Gaussian process formalism. \Cref{sec:inference-method} details the inference methodology, including the likelihood function, priors and emulator construction. We present inference on mock observations in \Cref{app:control} and on the observed intrinsic scatter in \Cref{sec:results}. We conclude in \Cref{sec:conclusions}. Throughout this paper we assume a flat $\Lambda$CDM cosmology with $H_0 = 70$ km/s/Mpc, $\Omega_m = 0.3$ and $\Omega_b = 0.05$. The analysis code used in this project is available at \url{https://github.com/htjb/SFV-SFMS}.

\begin{figure*}
    \centering
    \includegraphics[width=\linewidth]{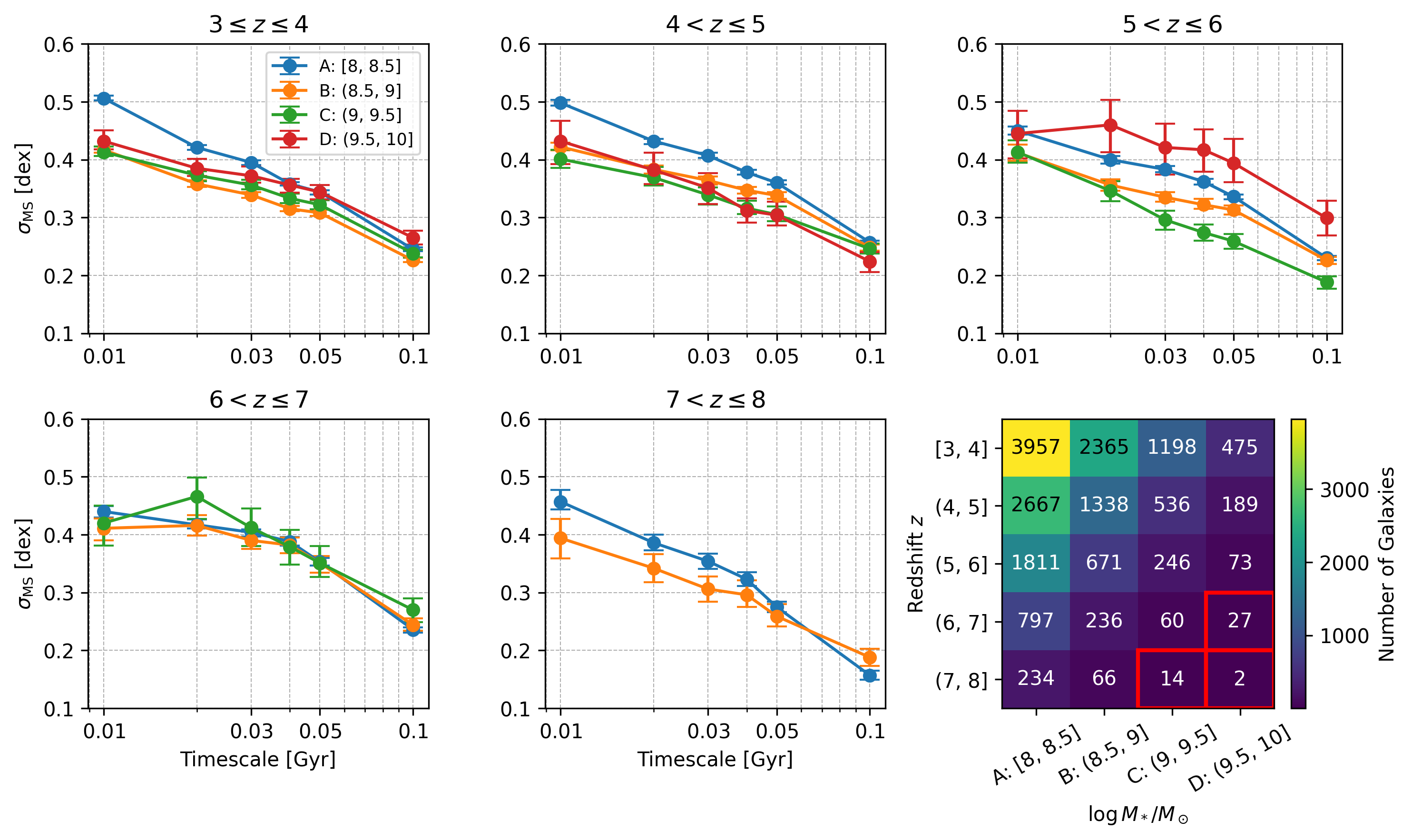}
    \caption{The intrinsic scatter around the main sequence on timescales between 10 and 100 Myr for redshifts between 3 and 8 (shown in panels) and stellar masses between $10^8$ and $10^{10}$ M$_\odot$ (indicated by colours). The error bars shown are estimated from bootstrapping and are later floored at 3\% in our analysis. The bottom right panel shows the number of galaxies in each stellar mass and redshift bin analysed in this work. Bins with fewer than 30 galaxies are excluded from our analysis and are highlighted in red. In total around 17,000 of the 48,156 galaxies in the catalogue analysed in \protect\cite{Simmonds2025burstingattheseams} are used to calculate the intrinsic scatters, shown in the other panels, around the main sequence after a stellar mass completeness cut  and redshift cut.}
    \label{fig:galaxy-numbers}
\end{figure*}

\section{Observational constraints on the SFMS}
\label{sec:observations}
The aim of this work is to forward-model the SFMS at $z \simeq 3-8$ as revealed by recent \textit{JWST} observations. Our focus is on the intrinsic scatter of the SFMS, quantified using SFRs averaged over different timescales, which provides direct constraints on the variability of star formation. Our analysis builds on the sample and methodology of \citet{Simmonds2025burstingattheseams}, who characterised the SFMS and its scatter to probe the burstiness of star formation across cosmic time. In brief, the dataset consists of a compilation of 48,156 galaxies at $3\leq z\leq 9$, with photometry from the JWST Advanced Deep Extragalactic Survey \citep[JADES; ][]{Eisenstein2023JADES,Bunker2024}, in the GOODS-N and GOODS-S \citep{Giavalisco2004} fields. All photometric measurements were extracted using Kron-convolved apertures, and a minimum error floor of 5\%, following \cite{Simmonds2025burstingattheseams}, was imposed to account for systematic uncertainties not captured by the reduction pipelines and for data-model mismatches. The 5\% floor prevents small formal photometric errors from dominating the spectral energy distribution (SED) fits discussed below when the model cannot reproduce every data point exactly. This mismatch between the data and model can reflect both photometric systematics and limitations in the stellar population synthesis; they are therefore not purely random measurement noise. The floor is implemented as a per-measurement uncertainty rather than an explicit model of a correlated systematics. Similar floors were explored in \cite{Mathews2023Emulators}.

The galaxy properties were derived with the SED modelling code \texttt{Prospector} \citep{Johnson2019,Johnson2021}. We refer the reader to \cite{Simmonds2024complete} and \cite{Simmonds2025burstingattheseams} for a detailed explanation of the fitting routine and the underlying assumptions adopted. However, of particular importance for this work, a non-parametric SFH was used \citep[continuity SFH prior;][]{Leja2019}, where the SFH is described by eight SFR bins. The minimum bin was set to a look-back time of 5 Myr, while the rest were divided equally in logarithmic space depending on the redshift. The ratios and amplitudes between adjacent SFR bins were allowed to vary following a Student's t-distribution with a width of 0.3, permitting (but not imposing) a modest degree of burstiness when the data favours it \citep{Tacchella2022}. The SFRs were then calculated at averaging timescales (t$_{\rm{avg}}$) of 10, 20, 30, 40, 50, and 100 Myr following: 
\begin{equation}
    \mathrm{SFR}_{t_\mathrm{avg}} = \frac{1}{t_\mathrm{avg}} \int^{t_\mathrm{avg}}_{0} \mathrm{SFR}(t^\prime) dt^\prime,
\label{eq:time-averaged-sfr}
\end{equation}
\noindent
where $t^\prime$ is the look-back time. The accuracy of the photometric redshifts was assessed by a comparison with spectroscopic redshifts where available \citep{Simmonds2024complete, Hainline2024}. Any uncertainty in the photometric-redshift propagates through the SED modelling code Prospector to the posterior estimates on stellar mass and SFR for each galaxy.

Once time averaged SFRs were estimated for each galaxy, \cite{Simmonds2025burstingattheseams} fit for the sSFR using
\begin{equation}
    \text{sSFR}_{\rm{MS}}(\text{M}_*,z)[\text{Gyr}^{-1}] = \text{s}_{\rm{b}}\times \left(\frac{\text{M}_*}{10^{10}\text{M}_{\odot}}\right)^{\beta}\times(1+z)^{\mu},
\end{equation}
following \cite{Tacchella2016}, where $\mu$ represents the redshift evolution of the SFMS, $\beta$ the stellar mass dependence, and s$_{\rm{b}}$ a general normalization in units of Gyr$^{-1}$. The fit was performed in the conservative parameter space where the sample is complete in stellar mass (log(M$_*$/M$_{\odot}$)$=9.0-10.3$), and was subsequently extrapolated down to log(M$_*$/M$_{\odot}$)$=8.0$.

In order to estimate the scatter of the SFMS as a function of redshift (up to $z=8$ due to low numbers at $z>8$), \cite{Simmonds2025burstingattheseams} divided their sample into four stellar mass bins:
\begin{itemize}
    \item A: $8.0\leq\log(\text{M}_*/\text{M}_{\odot})\leq 8.5$ 
    \item B: $8.5 <\log(\text{M}_*/\text{M}_{\odot})\leq 9.0$
    \item C: $9.0 <\log(\text{M}_*/\text{M}_{\odot})\leq 9.5$
    \item D: $9.5 <\log(\text{M}_*/\text{M}_{\odot})\leq 10.0$
\end{itemize}
The masses for each galaxy in the sample are inferred as part of the SED modelling. Each inferred mass has an associated error that means it might be binned incorrectly. Assessing this error would require posterior resampling or hierarchical modelling and this is left for future work.

The distance to the SFMS ($\Delta_{\rm{MS}}$) is calculated for each galaxy. Due to observational limitations, it is likely that galaxies with low SFRs (and therefore low $\Delta_{\rm{MS}}$) are missed \citep{Feldmann2017}. To circumvent this issue, the positive side of the $\Delta_{\rm{MS}}$ histograms were mirrored around zero \citep[proven to be a good estimation in][]{McClymont2025}. The observed scatter ($\sigma_{\rm{obs}}$) is defined as the standard deviation of the mirrored distributions. The intrinsic scatter ($\sigma_{\rm MS}$) is obtained by subtracting the uncertainties in stellar mass and SFR from the SED modelling in quadrature. For reference, the SED uncertainties, which include a contribution from the uncertainty in the photometric-redshift, vary between $\simeq0.1-0.3$ dex; they increase with shorter t$_{\rm{avg}}$ and as a function of redshift. Finally, the uncertainties in $\sigma_{\rm MS}$ were estimated by bootstrapping the distributions 100 times, and reflect the number of galaxies in each stellar mass bin. A minimum error of $3\%$ in the intrinsic scatter is assumed. The SFMS scatter $\sigma_{\rm MS}$ for the different stellar mass and redshift bins is shown in \cref{fig:galaxy-numbers} along with the number of galaxies in each stellar mass bin at redshifts between 3 and 8 (bottom right). Stellar mass and redshift bins with fewer than 30 galaxies are excluded from our analysis.

\section{Forward Modelling Variability of Star Formation}
\label{sec:simulations}

\begin{figure*}
    \centering
    \includegraphics[]{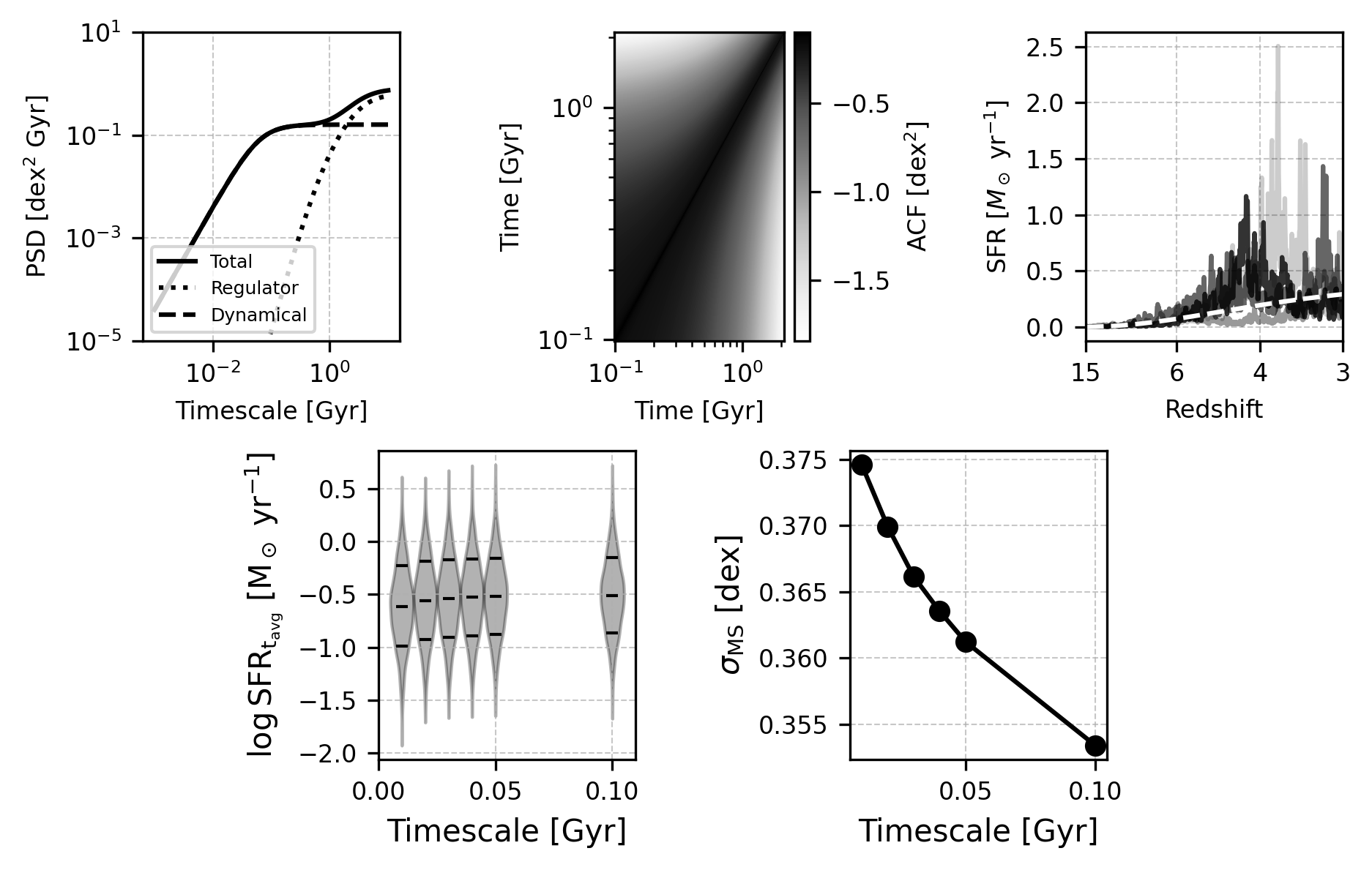}
    \caption{Example of the data generation pipeline described in \Cref{sec:simulations} using the \texttt{ExtReg} PSD \citep{Iyer2024ExtReg}. For each set of PSD parameters, describing the variance of the different PSD components and characteristic timescales of star formation, we calculate a corresponding power spectral density (top left) and auto correlation function (top middle). For the PSD we show the dynamical and regulator components as dashed and dotted lines respectively. The dynamical component dominates on short time scales for this particular PSD model. The ACF is shown as a heat map with the colour showing the strength of the correlation between the two times on the x and y axis e.g. $\tau=t' - t$. Using the ACF to define a Gaussian process, we can draw star-formation histories for a simulated population of galaxies (top right; different shades of grey show different draws). The PSD characterises the fluctuations around the mean SFR shown as the white dashed line. We integrate the star-formation histories over a range of different timescales to get corresponding distributions of $\log {\rm SFR}_{t_{\rm avg}}$ (bottom left). We show the mean and one sigma markers for the distributions. Taking the standard deviation of each distribution gives us $\sigma_{\rm MS}$ (bottom right).}
    \label{fig:example-data-gen}
\end{figure*}

In this section we describe our forward model, which generates simulated observations of $\sigma_{\rm MS}(\theta_{\rm PSD})$ to constrain the PSD parameters $\theta_{\rm PSD}$ describing the galaxy population in \cite{Simmonds2025burstingattheseams} and probe the processes driving their star formation. The forward model is illustrated in \cref{fig:example-data-gen}.

\subsection{Power Spectral Density}

The SFR of individual main sequence galaxies is assumed to oscillate around some redshift and stellar mass dependent average rate in a state of quasi-equilibrium with different astrophysical processes driving star formation on different time scales. The relative balance of these processes leads to bursts of star formation and periods of (mini-)quenching and the variablility can be modelled with a power spectral density (PSD).

A number of different models for the variability of star formation in main sequence galaxies have been proposed in the literature through different descriptions of the PSD. Phenomenological PSD models like the Simple Harmonic Oscillator~(\texttt{SHO}) try to capture the stochastic fluctuations in galaxy star-formation histories around the star-forming main sequence. The amplitude of the \texttt{SHO} model is given by
\begin{equation}
    S_\mathrm{\texttt{SHO}}(\omega) = \sqrt{\frac{2}{\pi}} \frac{S_0 \omega_0^4}{(\omega^2 - \omega_0^2)^2 + \frac{\omega_0^2 \omega^2}{Q^2}}
    \label{eq:SHO-psd}
\end{equation}
where $\omega = 2 \pi / \tau$, $S_0$ is the amplitude of the signal at $\omega_0 = 2\pi / \tau_0$ \citep{ForemanMackey2017SHO}. The \texttt{SHO} model is a flexible stochastic variability model that generalizes a damped random walk. $\tau_0$ defines the timescale for oscillations between periods of quenching and rapid star formation, and the parameter $Q$ defines how preferential this scale is for star formation. A high value of $Q$ essentially smooths fluctuations around the underlying sinusoidal SFHs predicted by the model.

An alternative formalism for the PSD is the regulator model. The regulator model captures fluctuations in the SFR of main sequence galaxies due to gas inflow and equilibrium of gas cycling between atomic and molecular hydrogen. The amplitude of the PSD is given by
\begin{equation}
    S_\mathrm{\texttt{Reg}}(\omega) =\ \frac{\sigma_\mathrm{\texttt{Reg}}^2}{
        1 + \left((2\pi \tau_\mathrm{\texttt{in}})^2 + (2\pi\tau_\mathrm{\texttt{eq}})^2\right)\omega^2 
        + (2\pi\tau_\mathrm{\texttt{in}})^2(2\pi\tau_\mathrm{\texttt{eq}})^2 \omega^4}
\end{equation}
where $\sigma_\mathrm{\texttt{Reg}}^2$ is the variance at $\omega=0$. $\tau_\mathrm{\texttt{in}}$ is the characteristic timescale for gas inflow to galaxies and $\tau_\mathrm{\texttt{eq}}$ captures how quickly the SFR of a galaxy reacts to the inflow of gas. Each component, inflow and equilibriation, are modelled as damped random walks and then coupled to each other.

In \cite{Tacchella2020ExtReg} the authors extended the regulator model to include an additional component describing internal dynamical processes that drive star formation such as star formation in spiral arms, bar instabilities and giant molecular cloud related physics. Since this is largely independent of gas inflow and cycling the power is additive and the amplitude of this extended regulator (\texttt{ExtReg}) model is given by
\begin{equation}
    S_\mathrm{\texttt{ExtReg}}(\omega) =\ S_\mathrm{\texttt{Reg}}(\omega)
    + \frac{\sigma_\mathrm{\texttt{Dyn}}^2}{1 + (2 \pi \tau_\mathrm{\texttt{Dyn}})^2 \omega^2}
\end{equation}
where $\sigma_\mathrm{\texttt{Dyn}}^2$ is the variance of dynamical process-driven star formation at $\omega=0$ and $\tau_\mathrm{\texttt{Dyn}}$ is the characteristic timescale for these dynamical processes. An example of the \texttt{ExtReg} power spectral density is shown in the top left panel of \cref{fig:example-data-gen}.

In this work we focus on constraining and comparing the \texttt{SHO} PSD and the \texttt{ExtReg} model, which encompasses the regulator model by definition. We note that the \texttt{SHO} is a single component model which in principle could capture either the dynamical processes of star formation or the longer timescale gas inflow/equilibriation driven star formation but not both like the two component \texttt{ExtReg} model.

\subsection{Gaussian Process for Star Formation}

The PSD for star formation can be related using the Wiener-Khinchin theorem to an Auto-Correlation Function (ACF) denoted $\mathcal{C}(\tau)$ which tells you how correlated the SFR at two different time scales is. The Wiener-Khinchin theorem states that
\begin{equation}
    \begin{aligned}
    S(\omega) = \int^{+\infty}_{-\infty} \mathcal{C}(\tau)  e^{-2\pi i \omega \tau} & d\tau \\
     & \Leftrightarrow \\
    \mathcal{C}(\tau) = & \int^{+\infty}_{-\infty} S(\omega) e^{-2\pi i \omega \tau} d\omega,
    \end{aligned}
\end{equation}
where $\tau = t^\prime - t$.

We can use the ACF to define a Gaussian process, which is the generalization of a Gaussian distribution to functional space, and from this we can draw SFRs  over time. Given a mean function $\mu(t) = \log \overline{\mathrm{SFR}}(t)$ and a covariance function $\mathcal{C}(\tau)$ we can define a Gaussian process for the SFR as
\begin{equation}
    \log \mathrm{SFR}(t) \sim \mathcal{N}(\log \mathrm{SFR}(t)|\log \overline{\mathrm{SFR}(t)}, \mathrm{C}(\tau)),
\end{equation}
where the mean function captures the general behaviour of the population of galaxies and the covariance paints fluctuations in the SFR on top of the mean behaviour.

For the \texttt{SHO} PSD the ACF is given by
\begin{equation}
    \mathrm{C}_\mathrm{\texttt{SHO}} (\tau) = S_0 \omega_0 Q e^{-\frac{\omega_0 \tau}{2 Q}} \bigg[\cos(\eta \omega_0\tau) + \frac{1}{2\eta Q} \sin(\eta \omega_0 \tau)\bigg],
\end{equation}
where $\eta = |1 - (4 Q^2)^{-1}|^{1/2}$ for $Q > 1/2$ \citep{ForemanMackey2017SHO} and the \texttt{ExtReg} ACF \citep{Iyer2024ExtReg} is given by
\begin{equation}
    \mathrm{C}_\mathrm{\texttt{ExtReg}} = \sigma_\mathrm{\texttt{Reg}}^2 \frac{\tau_\mathrm{\texttt{in}} e^{-|\tau|/\tau_\mathrm{\texttt{in}}} - \tau_\mathrm{\texttt{eq}} e^{-|\tau|/\tau_\mathrm{\texttt{eq}}}}{\tau_\mathrm{\texttt{in}} - \tau_\mathrm{\texttt{eq}}} + \sigma_{\rm \texttt{Dyn}}^2 e^{-|\tau|/\tau_\mathrm{\texttt{Dyn}}}.
\end{equation}
An example \texttt{ExtReg} ACF can be seen in the top middle panel of \cref{fig:example-data-gen} with $\sigma_{\rm \texttt{Reg}}= 0.8~, \tau_{\rm \texttt{eq}} = 0.5~{\rm Gyr}~, \tau_{\rm \texttt{in}}= 0.1~{\rm Gyr} ~, \sigma_{\rm \texttt{Dyn}} = 0.4~, \tau_{\rm \texttt{Dyn}} = 0.01~{\rm Gyr}$.

\subsection{The Secular Mean Star-Formation History}

We use a theoretical model, which gives rise to rising SFHs at early cosmic times and decreasing at later cosmic times, consistent with observed SFHs \cite[e.g.,][]{Simmonds2025burstingattheseams}, for the mean SFR given by $\overline{\rm SFR} (M_h, z) = f_b \epsilon_*(M_h) \dot{M_h}$ where $f_b$ is the baryon fraction, $\epsilon_*$ is the star-formation efficiency as a function of halo mass $M_h$ and $\dot{M_h}$ is the dark matter halo accretion rate \citep{Tacchella2018}.

We use a double power law model for the star-formation efficiency as a function halo mass
\begin{equation}
    \epsilon_\star (M_h) = \frac{2 \epsilon_0}{(M_h/M_0)^{- \alpha} + (M_h/M_0)^\beta}
    \label{eq:star-formation-efficiency}
\end{equation}
where $M_0 = 10^{12}$ M$_\odot$, $\epsilon_0 = 0.1$, $\alpha = 0.6$ and $\beta=0.5$ \citep{Shen2023impact}. The dark matter accretion rate is given by
\begin{equation}
    \begin{aligned}
        & \dot{M_h} = ~C \bigg(\frac{M_h}{10^{12}{\rm M}_\odot}\bigg)^\gamma \frac{H(z)}{H_0} \\
        & \gamma = 1.000 + 0.329 a - 0.206 a^2 \\
        & \log_{10} C = 2.630 - 1.828 a + 0.654 a^2
    \end{aligned}
    \label{eq:accretion-rate}
\end{equation}
where $a = 1/(1 + z)$ is the scale factor \citep{Rodriguez-Puebla2016}.

We integrate the accretion rate to get the halo mass history $M_h(z)$, assuming some initial seed halo mass at $z=30$, then for each $M_h$ and $z$ evaluate the mean ${\rm SFR}$. We are assuming each galaxy in our population has the same average halo mass history, and the choice of seed halo mass determines the mean stellar mass for galaxies observed at a particular redshift as shown in \cref{fig:seed-mass-stellar-mass}. In practice there is likely to be some galaxy-to-galaxy variation in the halo mass accretion histories and this would introduce additional scatter in the populations SFRs. However, these differences would be on timescales comparable to the halo dynamical time and would contribute an approximately $t_{\rm avg}$-independent component to $\sigma_{\rm MS}$ over 10-100 Myr and consequently this variation is unlikely to be probed by our data. A full investigation of this effect is left for future work.

An example of the mean function for galaxies at $z=3$ and a seed halo mass of $\log M_h^{\rm seed} / M_\odot = 7$ at $z=30$ is shown as a white dashed line in \cref{fig:example-data-gen} with several star-formation histories in different shades of gray drawn from the Gaussian process corresponding to the \texttt{ExtReg} PSD in the top left panel. The mean SFR peaks around cosmic noon and the SFRs vary from $\sim 0.1$ to $\sim 2.5$ solar masses per year.

\subsection{Intrinsic Scatter of Star Formation}

Using the discussed Gaussian process we generate simulated galaxy catalogues and for each estimate the intrinsic scatter $\sigma_{\rm MS}$ for star formation for a given set of PSD parameters $\theta_\mathrm{PSD}$. We apply the same selection cuts that were applied to the observations generating different data sets for each stellar mass and redshift bin. For a given set of PSD parameters, redshift bin and stellar mass bin we draw $N$ star-formation histories from the Gaussian process. To ensure that we generate catalogues of galaxies with stellar masses in the range of the mass bins at the relevant redshifts, the seed halo mass of the mean function is determined by interpolating the relationship in \cref{fig:seed-mass-stellar-mass}. Galaxies that fall outside the stellar mass bin range are discarded and if fewer than 30 galaxies survive the selection cut the PSD sample is also discarded.

For PSD parameters that produce more than 30 galaxies in the stellar mass bin, we then integrate each SFR in the mock catalogue over the same series of averaging time scales used to generate the data (bottom left in \cref{fig:example-data-gen}). For each timescale, we calculate the standard deviation of all $N$ ${\rm SFR}_{t_{\rm avg}}$ to get an estimate of the intrinsic scatter $\sigma_{\rm MS}$  (bottom right in \cref{fig:example-data-gen}) to compare with the observations in \cite{Simmonds2025burstingattheseams}. We note that the mean of the distribution of $\log {\rm SFR}_{t_{\rm avg}}$ changes on different averaging timescales and that this change varies with the parameters of the PSD. This is consistent with what is seen in \cite{Caplar2019} and the variation in the mean of the distribution could also be used to constrain $\theta_{\rm PSD}$. Exploration of this is, however, left for future work.

\section{Inference Pipeline, Emulation and Mock Data}
\label{sec:inference-method}

\subsection{Bayesian Inference}

We use nested sampling to infer the PSD parameters for the different observed intrinsic scatters. The posterior probability, which tells us how probable one set of parameters $\theta_{\rm PSD}$ are given our observed intrinsic scatters $D$ and forward model $M$ described in \cref{sec:simulations}, is given by
\begin{equation}
    P(\theta_\mathrm{PSD}|D, M) = \frac{P(D|\theta_\mathrm{PSD}, M)P(\theta_\mathrm{PSD}|M)}{P(D|M)}.
\end{equation}
$\mathcal{L}(D) = P(D|\theta_{\rm PSD}, M)$ is known as the likelihood and tells us how likely our data is given our parameters and model. The prior $\pi(\theta_{\rm PSD}) = P(\theta_{\rm PSD}|M)$ encodes our prior knowledge about the parameters of our model and finally the Bayesian evidence $\mathcal{Z} = P(D|M)$ encodes the probability of the data given our modelling choices. The likelihood and prior are both defined by the user and nested sampling calculates the evidence and returns samples from the posterior distribution. The evidence allows us to compare the different PSD models. See \cite{Skilling2006NS} and \cite{Ashton2022NS} for more details on how nested sampling works and is used in the physical sciences.

We assume that any error $\sigma(t_{\rm avg})$ in the measurements of $\sigma_{\rm MS}$, after correcting for uncertainties introduced by SED modelling, comes from for example catalogue incompleteness and is Gaussian in nature. Since the likelihood should follow the distribution of the random component of our data after subtracting our model it is given by
\begin{equation}
    \begin{aligned}
        \log \mathcal{L}(D) = -\frac{N}{2} & \log(2\pi \sigma(t_\mathrm{avg})^2) - \\ & \sum_i^N \frac{1}{2}\frac{(\sigma_{\rm MS}(t_\mathrm{avg}) - M(t_\mathrm{avg}|\theta_\mathrm{PSD}))^2}{\sigma(t_\mathrm{avg})^2}.
    \end{aligned}
\end{equation}
We use the JAX-based implementation of nested sampling in the \textsc{blackjax} package \citep{Cabezas2024blackjax, Yallup2026NSS}. The priors on our PSD parameters are outlined in \cref{tab:priors} and are uniform or log-uniform distributions representing our initial assumptions about the plausible values of the parameters.

\begin{table}
    \centering
    \begin{tabular}{|c|c|}
        \hline
        Parameter & Prior\\
        \hline
        \multicolumn{2}{|c|}{\texttt{SHO}} \\
        \hline
         $\log_{10}(S_0)$ &  $\mathcal{U}(-4, \log_{10}(5))$\\
         $\log_{10}(\tau_0$ {\rm[Gyr]}) &  $\mathcal{U}(-3, \log_{10}(3))$ \\
         $\log_{10}(Q)$ & $\mathcal{U}(\log_{10}(0.51), \log_{10}(5))$\\
         \hline
         \multicolumn{2}{|c|}{\texttt{ExtReg}} \\
         \hline
         $\log_{10}(\sigma_\mathrm{\texttt{Reg}})$ &  $\mathcal{U}(-2, 1)$\\
         $\log_{10}(\tau_\mathrm{\texttt{eq}} {\rm[Gyr]})$ & $\mathcal{U}(-3, 1)$\\
         $\log_{10}(\tau_\mathrm{\texttt{in}} {\rm[Gyr]})$ &  $\mathcal{U}(-2, 0)$\\
         $\log_{10}(\sigma_\mathrm{\texttt{Dyn}})$ & $\mathcal{U}(-2, \log_{10}(5))$\\
         $\log_{10}(\tau_\mathrm{\texttt{Dyn}} {\rm[Gyr]})$ & $\mathcal{U}(-3, -1)$\\
         \hline
         
    \end{tabular}
    \caption{The prior on the PSD parameters adopted in this paper.}
    \label{tab:priors}
\end{table}

\subsection{Emulation}

Calculating the covariance matrix for the Gaussian process over a fixed period at sufficiently fine enough resolution to capture important variations in the SFR for each galaxy is computationally expensive (even with modern frameworks like JAX). We therefore train neural network emulators to predict $\sigma_\mathrm{MS} (\theta_\mathrm{PSD})$ using a modified version\footnote{We use version 0.4.0 of the \textsc{astroemu} code (\url{https://github.com/htjb/astroemu}) with custom data loaders that can handle in memory arrays.} of the JAX-based framework \texttt{astroemu}.

The intrinsic scatter of star formation in a population of galaxies is directly related to the stellar masses of the galaxies and our simulation pipeline is designed to give us galaxy populations that fall within the same stellar mass and redshift bins as the observations. There are five redshift bins, four stellar mass bins and two PSD models explored in this work. For each combination we generate 2,500 simulated populations of 10,000 galaxies with different PSD parameters. As discussed, we use the relationship in \cref{fig:seed-mass-stellar-mass} to pick an appropriate halo seed halo mass at redshift 30 to create a population of galaxies that fall within the stellar mass bin at the observed redshift. We are assuming that the mean SFH is the same for all galaxies in a given stellar mass-redshift bin, and that all the galaxies in our simulated surveys form in halos with the same seed halo mass at $z=30$. We generate 34 different sets of training data (for the different combinations of PSD, stellar mass bin and redshift bin) and for each we have between roughly 2000 and 2500 $\sigma_{MS}(\theta_{PSD})$ to train our emulators, after applying the same cuts applied to the data. The data generation pipeline prevents the erroneous inclusion of low or high stellar mass galaxies in our training data and ensures that the scatters are representative of the observational datasets.

\begin{table}
    \centering
    \begin{tabular}{ll}
        \hline
        \textbf{Hyperparameter} & \textbf{Search Space} \\
        \hline 
        \texttt{Hidden Layers} & Integer in [1, 10] \\
        \texttt{Hidden Nodes} & Integer in [$N_{\theta_\mathrm{PSD}} + 1$, 32] \\
        \texttt{Learning Rate} & Log-uniform in [$10^{-4}$, $10^{-1}$] \\
        \texttt{Weight Decay} & Log-uniform in [$10^{-6}$, $10^{-2}$] \\
        \texttt{Batch Size} & \{32, 64, 128, 256, 512\} \\
        \texttt{Activation} & \{ReLU, LeakyReLU, Tanh, GELU\} \\
        \texttt{Log} $\theta_{\rm PSD}$ & \{True, False\} \\
        \hline
    \end{tabular}
    \caption{We optimize the hyperparameters used to train the neural network emulators with the \texttt{optuna} python package. The table outlines the search space.}
    \label{tab:emulator-hyperparameter-optimization}
\end{table}

\begin{figure}
    \centering
    \includegraphics[width=\linewidth]{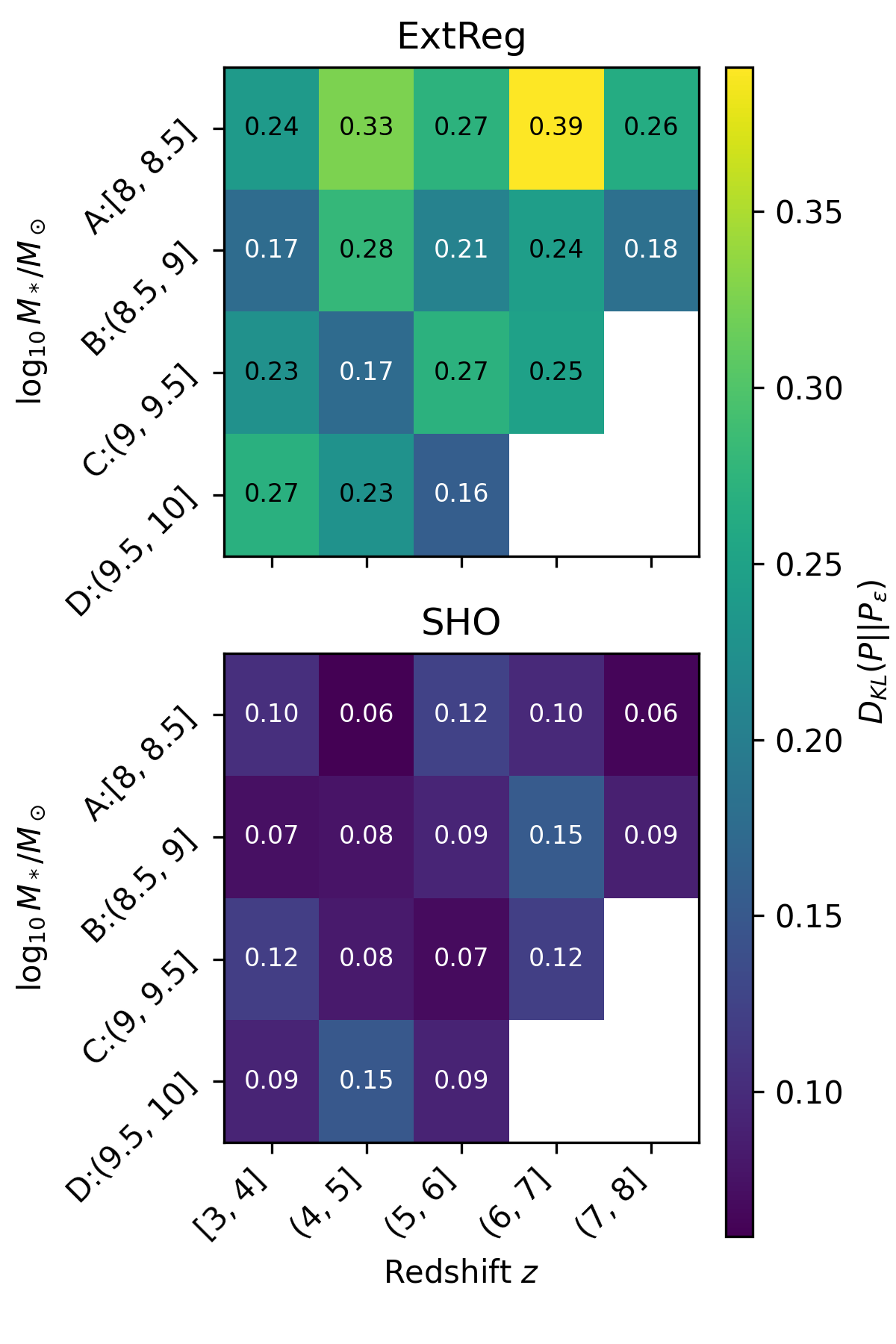}
    \caption{The upper bound on the incorrect information inferred when performing inference with each of the 34 emulators trained in this paper in nats. The top panel corresponds to the \texttt{ExtReg} emulators and the bottom to the \texttt{SHO} emulators. The \texttt{ExtReg} is a more complex model and the forecast uncertainties in the emulated posteriors are higher for this model than for the \texttt{SHO} model as a result. To account for the emulator error we include an additional uncertainty term in our likelihood.}
    \label{fig:dkl-errors}
\end{figure}

We use a fully connected feed forward neural network to emulate the relationship between $\sigma_\mathrm{MS}(t_\mathrm{avg})$ and $\theta_{\rm PSD}$. The architecture consists of an input layer, a configurable number of hidden layers with non-linear activations, an output layer and skip connections every two layers which stabilize training. The networks use the same tiling trick outlined in \cite{bevins2021globalemu} taking $\theta_{\rm PSD}$ and $t_{\rm avg}$ as inputs outputting a single value of $\sigma_\mathrm{MS}$ at the given $t_\mathrm{avg}$. A vectorised call is then used to emulate the intrinsic scatter at the averaging timescales of interest. The tiling trick reduces the number of outputs from six to one at the cost of having one more input to the network and has several advantages over learning the full spectrum or a PCA decomposition including: the target function is simpler to learn, the training data is augmented by a factor of $N_{t_{\rm avg}}$ and it ensures that $\sigma_{\rm MS}$ is learnt as a smooth continuous function of $t_{\rm avg}$.

The accuracy of the trained neural network will depend on the hyperparameters of the network; the number of layers, size of each layer, the optimizer used to fit the parameters of the network, the optimizers settings and the activation function. For each of the 34 emulators, we optimize the hyperparameters of the neural network using the python package \texttt{optuna} \citep{Akiba2019Optuna}. We run 100 trials, with for example different architectures, using the default Tree-structured Parzen Estimation algorithm \citep{Watanabe2023TPE} with early stopping. The hyperparameter search space is outlined in \cref{tab:emulator-hyperparameter-optimization}. Specifically, we optimize over the architecture of the network, the learning rate, the strength of the weight decay in the AdamW optimizer \citep{Loshchilov2017adamw}, the batch size, activation and whether the input parameters $\theta_{\rm PSD}$ should be in log or linear space. By default, we input $\log_{10}(t_{\rm avg})$ and predict $\log_{10} \sigma_{\rm MS} (t_{\rm avg})$. We then also standardise (by subtracting the mean and dividing by the standard deviation of the examples) the training, test and validation data sets which are comprised of $\approx 66\%$,$\approx 17\%$ and $\approx 17\%$  of the simulated data respectively. We train using a mean squared error loss
\begin{equation}
    L = \frac{1}{N} \sum_i (\sigma_{{\rm MS,} i}^{\rm true}(t_{\rm avg}) - \sigma_{{\rm MS,} i}^{\rm pred}(t_{\rm avg}))^2,
\end{equation}
and optimise our hyperparameter choices to minimise the average RMSE across the test data $\epsilon = \sqrt{L}$.

When using an emulator for inference the recovered posterior $P_\epsilon(\theta_{\rm PSD}| D, M_\epsilon)$ is an approximation to the true underlying posterior $P(\theta_{\rm PSD}| D, M)$ since the output of the emulator $M_\epsilon$ is only an approximation to the true model $M$ which generated the training data. \cite{Bevins2025Accuracy} outlined an approximate relationship for the Kullback-Leibler (KL) divergence between these two distributions. The KL divergence is an information theoretic metric that quantifies the difference between two distributions. The relationship is dependent on the magnitude of the noise in the data $\sigma$ and the emulator error $\epsilon$
\begin{equation}
    \mathcal{D}(P||P_\epsilon) \leq \frac{N_d}{2}\bigg(\frac{\epsilon}{\sigma}\bigg)^2,
    \label{eq:information_loss}
\end{equation}
where $\sigma$ and $\epsilon$ are in the same units and $N_d$ is the number of data points. While $\epsilon$ tells us how inaccurate our predictions of $\sigma_{\rm MS}$ are the \cref{eq:information_loss} gives us an upper bound on how accurate our posteriors on the parameters $\theta_{\rm PSD}$ will be. If our emulator was a perfect representation of the underlying model then $\mathcal{D}(P||P_\epsilon) = 0$ and the authors of \cite{Bevins2025Accuracy} suggest a $D_{\rm KL}(P||P_\epsilon) \lesssim 1$ corresponds to a high degree of accuracy. We report the upper bounds on the KL divergence in \cref{fig:dkl-errors} using the average error on the data in the relevant stellar mass bin. We find that the emulator error is generally larger for the \texttt{ExtReg} model but that the predicted upper bound on the KL divergence between the true and emulated posterior is $<0.4$ nats when using the mean fractional uncertainty on the emulated test data $\bar{\epsilon}$.

To account for uncertainty introduced by the emulator, we include an additional term in our likelihood function so that it becomes
\begin{equation}
\begin{aligned}
    \sigma = & \sqrt{\sigma(t_\mathrm{avg})^2 + (\bar{\epsilon}~~\sigma_{\rm MS}(t_\mathrm{avg}))^2},\\
    \log \mathcal{L}(D) = & -\frac{N}{2} \log(2\pi \sigma^2) \\ & - \sum_i^N \frac{1}{2}\frac{(\sigma_{\rm MS}(t_\mathrm{avg}) - M(t_\mathrm{avg}|\theta_\mathrm{PSD}))^2}{\sigma^2}.
\end{aligned}
\end{equation}
This is a common practice with $\bar{\epsilon} \sigma_{\rm MS}$ effectively representing our uncertainty in our model \citep[e.g. see][]{GesseyJones2024SuperStrings, Pochinda2024PopIII, Dhandha2025JWST21cm}.

\subsection{Validation with Mock Data}
\label{app:control}

\begin{figure*}
    \centering
    \includegraphics[width=0.45\linewidth]{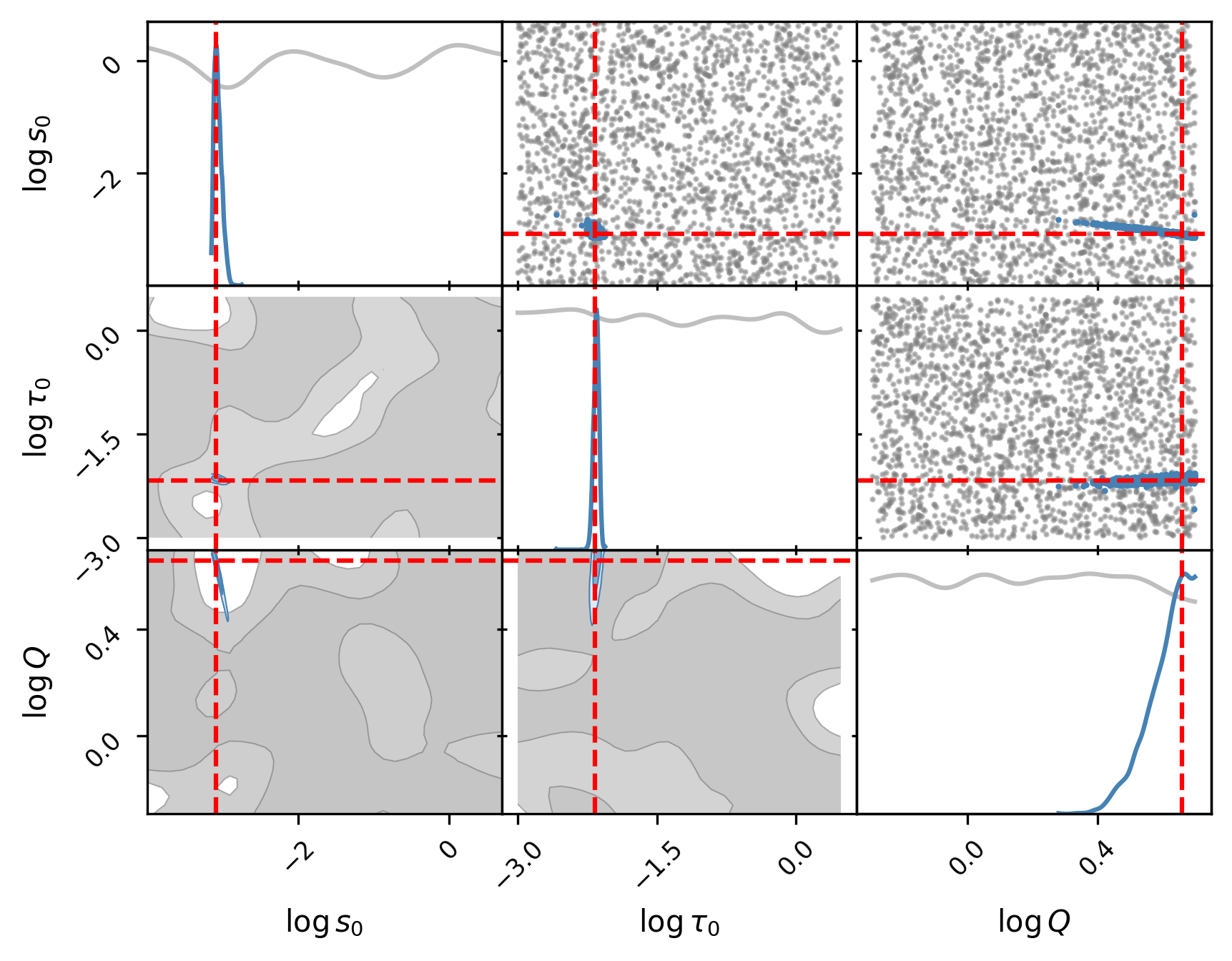}
    \includegraphics[width=0.45\linewidth]{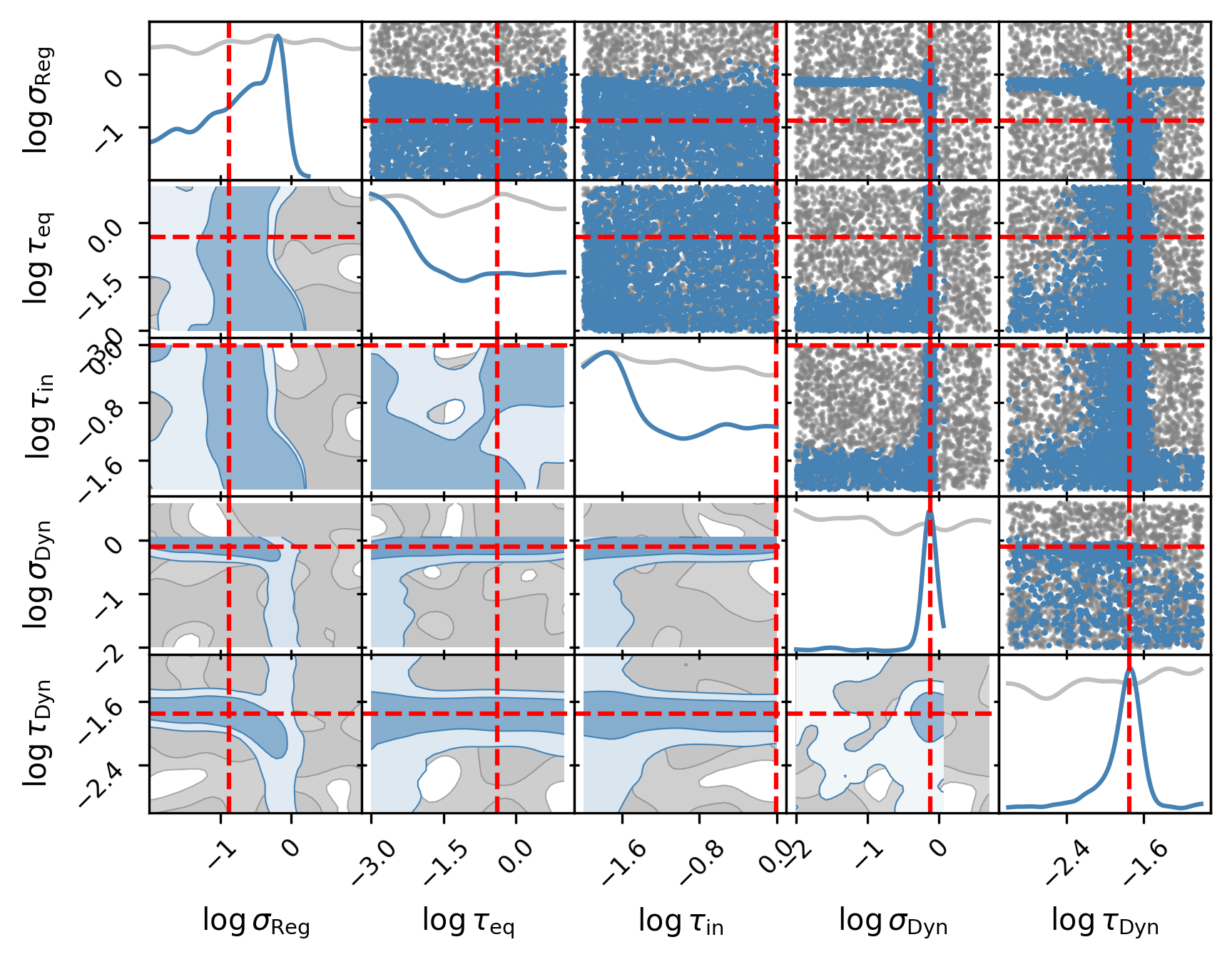}
    \caption{Parameter recovery for mock data in the stellar mass bin $\log M_*/M_\odot = 8 - 8.5$ and redshift $z=4-5$ emulator for the \texttt{SHO} model (left) and \texttt{ExtReg} model (right). The prior is shown in grey, the posterior in blue and the true values as red dashed lines.}
    \label{fig:control_hist}
\end{figure*}

\begin{figure*}
    \centering
    \includegraphics[width=0.43\linewidth]{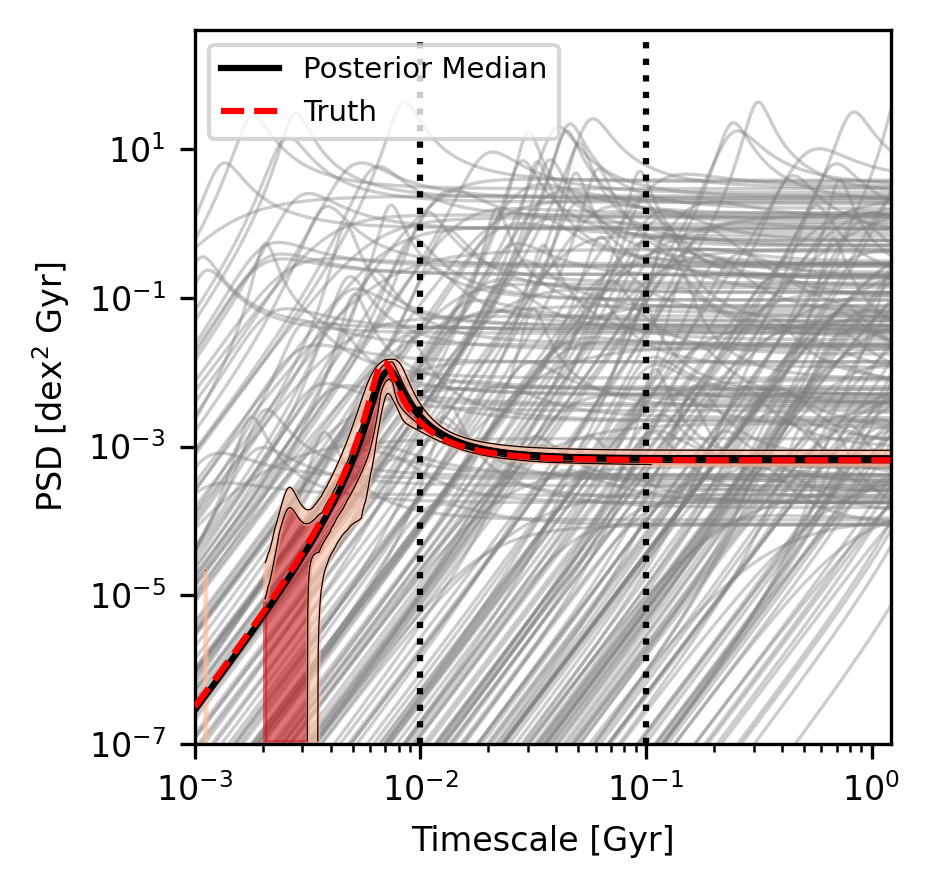}
    \includegraphics[width=0.43\linewidth]{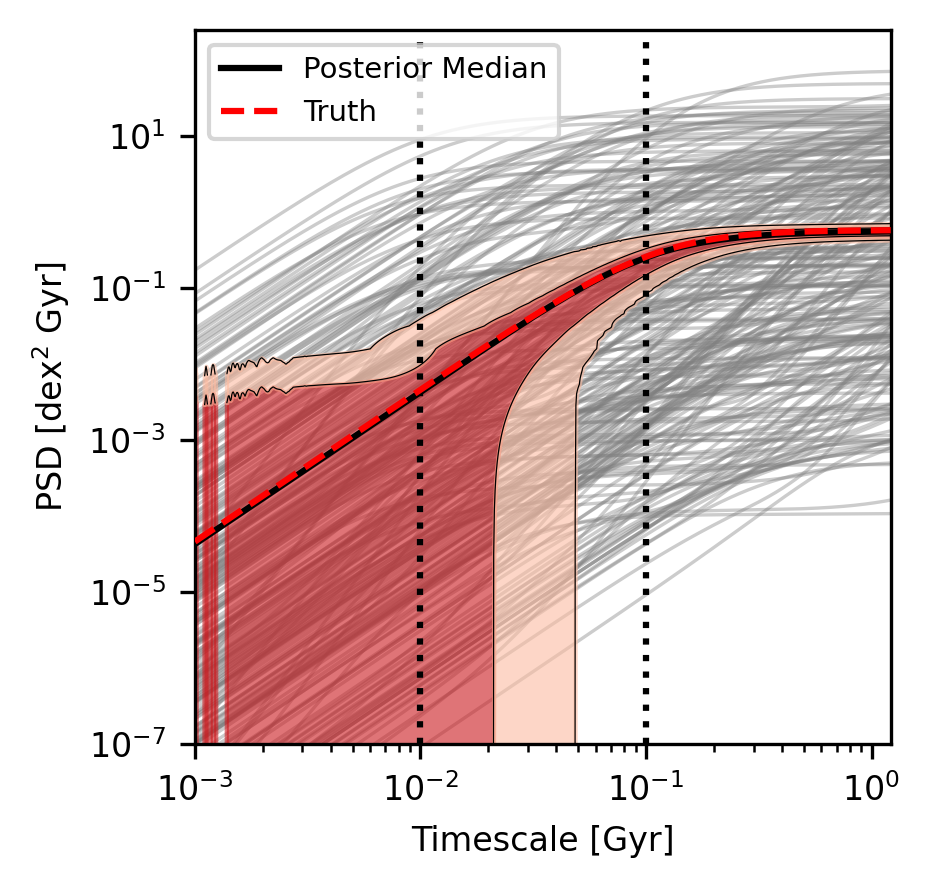}
    \caption{The posteriors on the PSDs for mock observations in the $\log M_* / M_\odot = 8 - 8.5$ and redshift $z = 4 - 5$ bin for the \texttt{SHO} model (left) and \texttt{ExtReg} model (right). Prior samples are shown as grey lines, one and two sigma posterior contours are shown as dark and light shaded red areas, the average of the PSD posterior is shown as the black solid line and the true PSD is shown as the red dashed line. The two vertical dotted lines show timescales of 10 and 100 Myr.}
    \label{fig:control_psd}
\end{figure*}

\begin{figure*}
    \centering
    \includegraphics[width=0.45\linewidth]{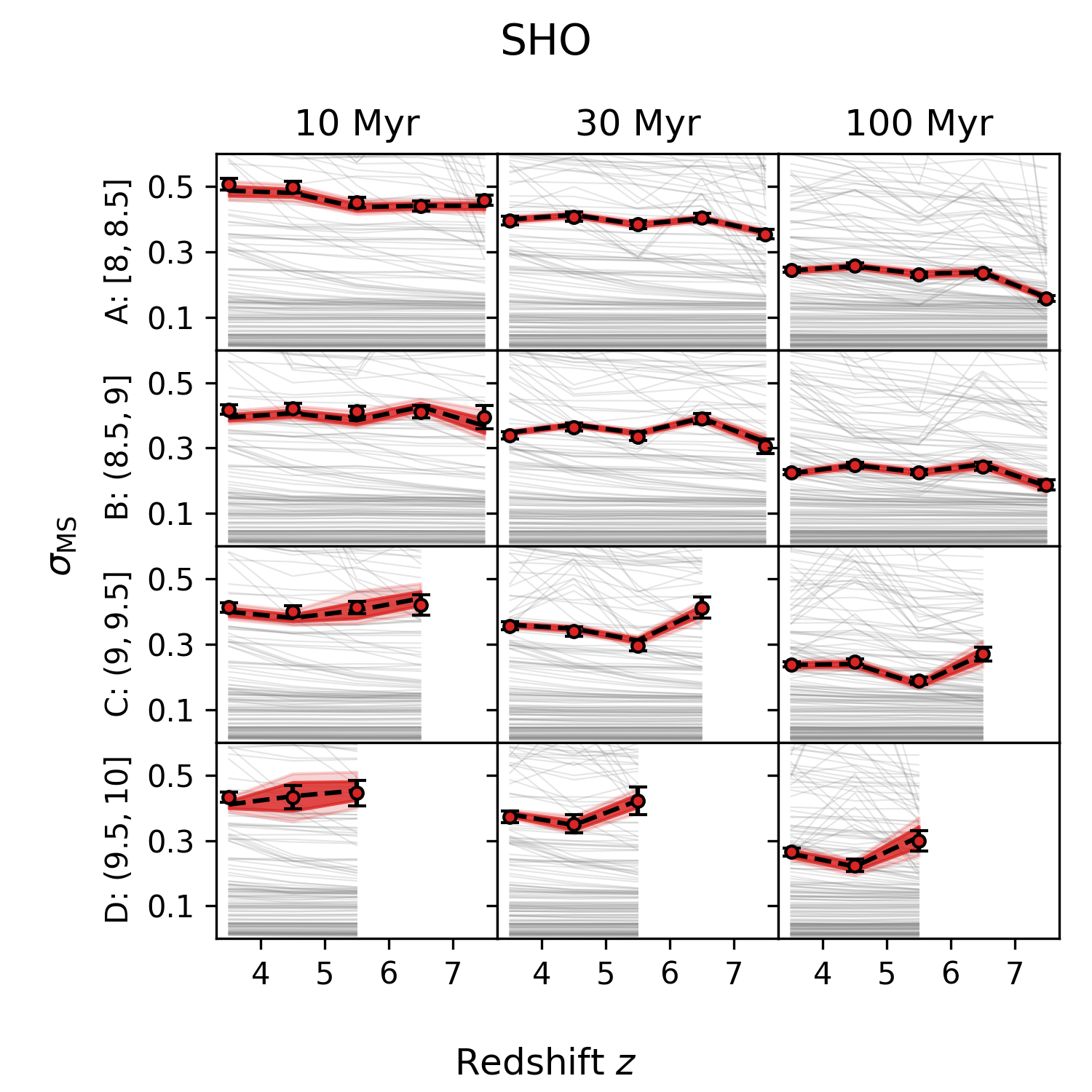}
    \includegraphics[width=0.45\linewidth]{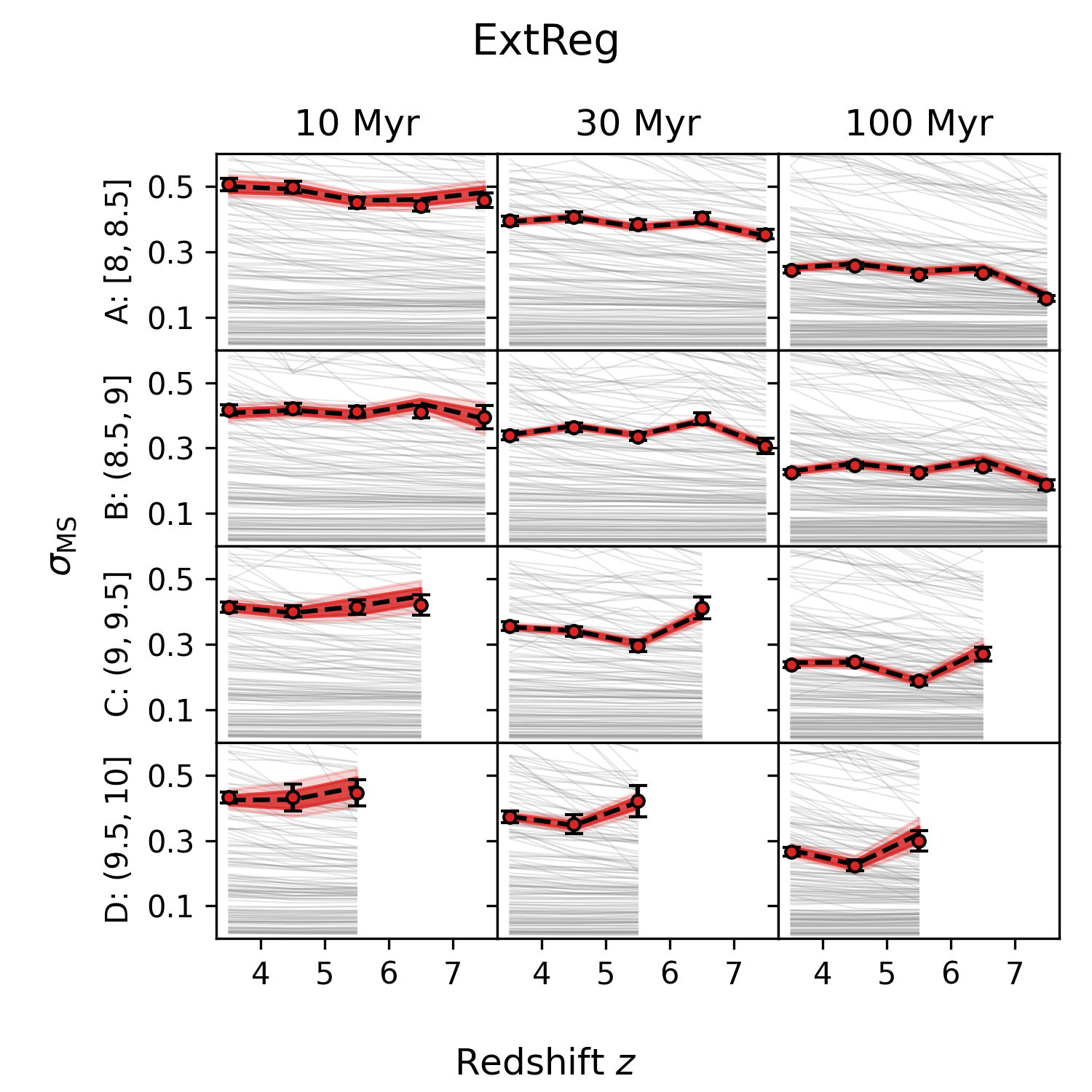}
    \caption{Predicted main sequence scatter ($\sigma_{\rm MS}$) as a function of averaging timescale for the \texttt{SHO} model (\textit{left}) and \texttt{ExtReg} model (\textit{right}). Each row shows results for one stellar mass bin, at three different averaging timescales over the redshift range 3-8. Dashed black lines show the average of the functional posteriors, while shaded regions indicate 68\% (dark) and 95\% (light) posterior confidence intervals. Prior samples are shown in grey. Data points with error bars show the observed scatter from \protect\cite{Simmonds2025burstingattheseams}. Note that we are fitting the data as a function of time at a given mass and redshift bin and not as a function of redshift as plotted. We are linearly interpolating between the mean posterior values and one and two sigma contour limits in each panel.}
    \label{fig:sigma_confidence}
\end{figure*}

To validate our pipeline, we perform inference on mock observations generated from known input PSD parameters. These tests ensure that: the neural network emulators accurately reproduce the forward model and the likelihood function and nested sampling correctly recover input parameters. For each PSD model, redshift bin and stellar mass bin, we randomly select three example intrinsic scatters from the emulator test data and assign an error equal to the mean fractional error on the test data predictions from the emulator or 3\% (whichever is larger). For each data set we run the full inference pipeline to recover posterior distributions. We show the results for inference on one mock observation for galaxies in the stellar mass bin $\log M_* / M_\odot = 8 - 8.5$ and redshift bin $z = 4-5$ in \cref{fig:control_hist} and \cref{fig:control_psd}. We accurately recover the parameters of the \texttt{SHO} kernel, the parameters of the dynamical component of the \texttt{ExtReg} component and the PSDs. While we are unable to accurately recover the true values of the parameters of the regulator component of the \texttt{ExtReg} model they are well within the posterior. 

Additional results for other stellar mass and redshift bins can be found with the inference chains on Zenodo and the code can be found on the github. In general, we find that we can recover $\sigma_{\rm reg}$ and $\sigma_{\rm dyn}$ well and $\tau_0$ and $\tau_{\rm dyn}$ are both well constrained provided that the true timescales fall within or below the 10-100 Myr period probed by the data. Where we are unable to recover the parameters accurately (e.g. $\tau_{\rm eq}$, $\tau_{\rm in}$ and $\sigma_{\rm reg}$ in \cref{fig:control_hist}) the true values typically fall within the posterior. Across the 51 trials for each of the \texttt{SHO} and the \texttt{ExtReg} models, we find that the true parameter value lies within the 68\% confidence interval of the posterior in 57-73\% of the cases and in the 95\% confidence interval 90-98\% of the time depending on the parameter.

\section{Results}
\label{sec:results}

\begin{figure*}
    \centering
    \includegraphics[width=\linewidth]{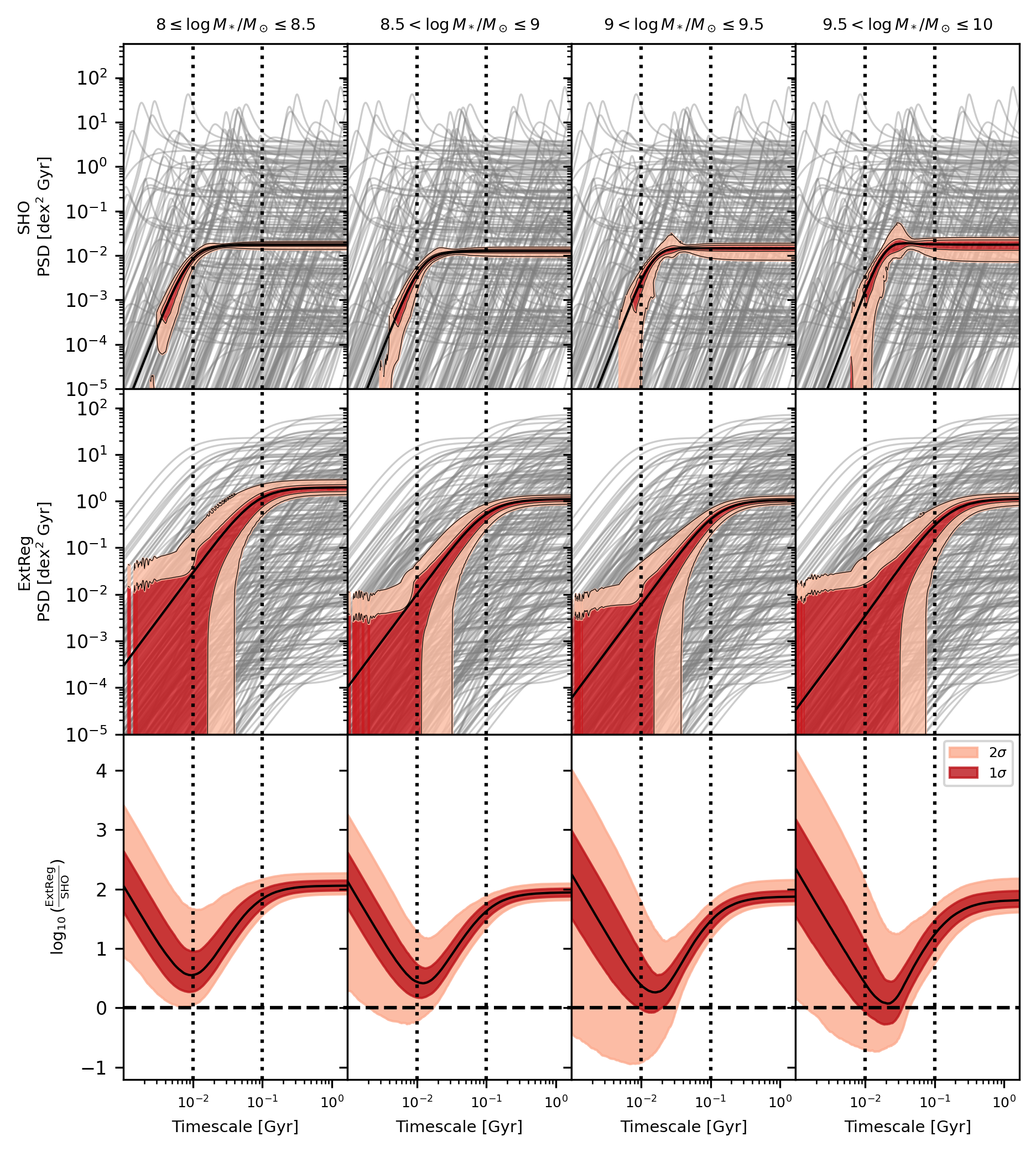}
    \caption{The posterior distributions on the PSD constraints for all four stellar mass bins at redshifts $3-4$. The top row corresponds to the \texttt{SHO} PSD, the middle row to the \texttt{ExtReg} model and the bottom row to the log of the ratio between them. The vertical dashed lines in all panels correspond to timescales of 10 and 100 Myr. Prior samples are shown as grey lines in the top two rows. One sigma and two sigma confidence regions are shown as dark and light shaded regions and the solid black lines are the functional averages of the PSD posteriors and differences. The PSD posteriors are most consistent on timescales of 10-50 Myr but differ on long and short timescales. The maximum timescale plotted corresponds to the lookback time from $z=3.5$ to $z=30$. PSD posteriors are visually consistent across stellar mass bins indicating a lack of evolution in star formation with stellar mass at $z=3-4$ for masses $\log M_* / M_\odot = 8 -10$.}
    \label{fig:psd-plots}
\end{figure*}

In the following sections we look at the functional constraints from the real data on both the PSD and $\sigma_{\rm MS}$, if there is any preference for one model over the other, and the implications of our analysis on star formation between redshift 3-8.

\subsection{The Power Spectral Density and Main Sequence Scatter}

Both the \texttt{ExtReg} and \texttt{SHO} model are able to model the main sequence scatters across redshift and stellar mass well as can be seen in \cref{fig:sigma_confidence}. We show the one and two $\sigma$ contours on $\sigma_{MS}(\theta_{\rm PSD})$ at three different time scales 10, 30 and 100 Myr for the \texttt{SHO} model on the left and \texttt{ExtReg} model on the right. The models are able to capture the general decrease in scatter (burstiness) with time and the uncertainty in the data is well reflected in the posteriors.

The close agreement between the data and the models suggests that the PSDs should also agree on timescales between 10 and 100 Myr. \cref{fig:psd-plots} shows the posterior distribution for the PSDs in the $z = 3 -4$ bin and all four stellar mass bins for both the \texttt{SHO} model (top row) and the \texttt{ExtReg} model (middle row). The dark shaded region shows the $1\sigma$ constraints and the light shaded region shows the $2\sigma$ constraints on the PSDs. Samples from the prior are shown as grey lines. The bottom row shows the log of the ratio between the \texttt{ExtReg} PSD and the \texttt{SHO} PSD posteriors. The PSDs agree most closely on timescales of 10-50 Myr with the largest discrepencies on timescales $\gtrsim 100$ Myr and $\lesssim 10$ Myr. The vertical dotted lines in all panels show 10 Myr and 100 Myr timescales.

\begin{figure*}
    \centering
    \includegraphics[width=\linewidth]{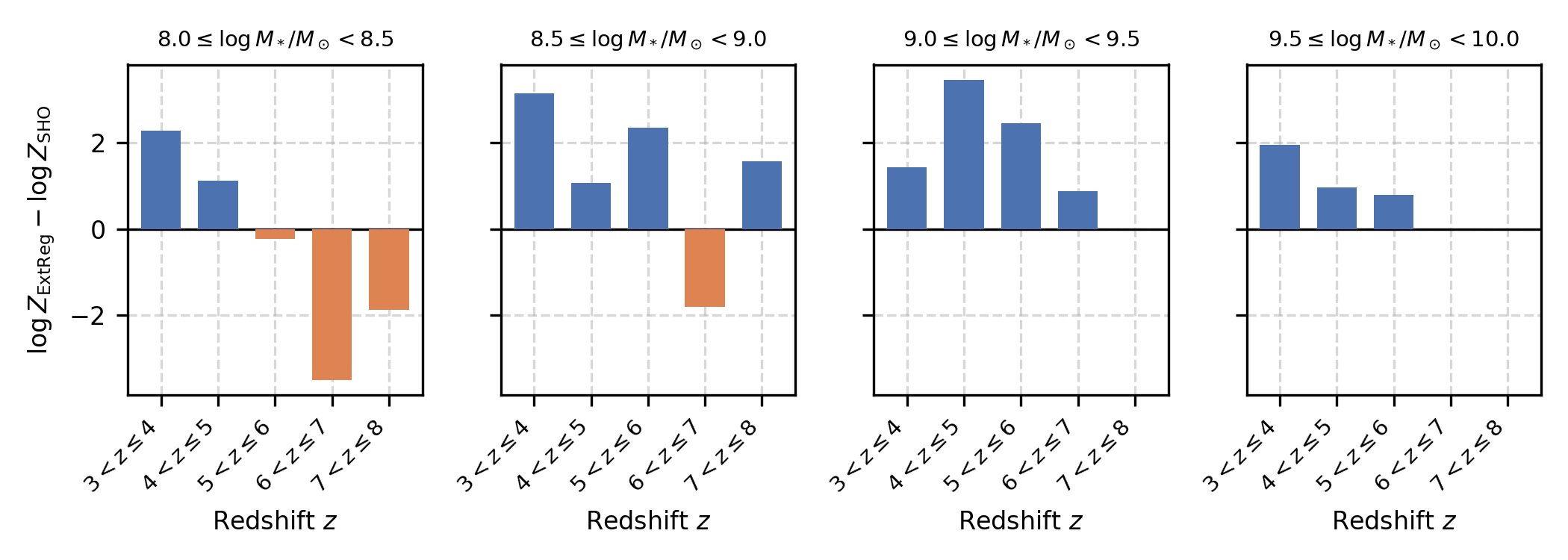}
    \caption{Bayesian model comparison between the \texttt{ExtReg} and \texttt{SHO} models across the different stellar mass and redshift bins. Logarithmic evidence differences $\Delta \log \mathcal{Z} = \log \mathcal{Z}_{\rm \texttt{ExtReg}} - \log \mathcal{Z}_{\rm \texttt{SHO}}$ are shown with positive values indicating a preference for the \texttt{ExtReg} model and negative values for the \texttt{SHO} model. A log Bayes ratio of $3$ nats (equivalently $\approx 1.3$ in base 10) corresponds to betting odds of 20:1 and is regarded as strong evidence on a Jefferys scale.}
    \label{fig:model_comparison}
\end{figure*}

\begin{figure*}
    \centering
    \includegraphics[width=\linewidth]{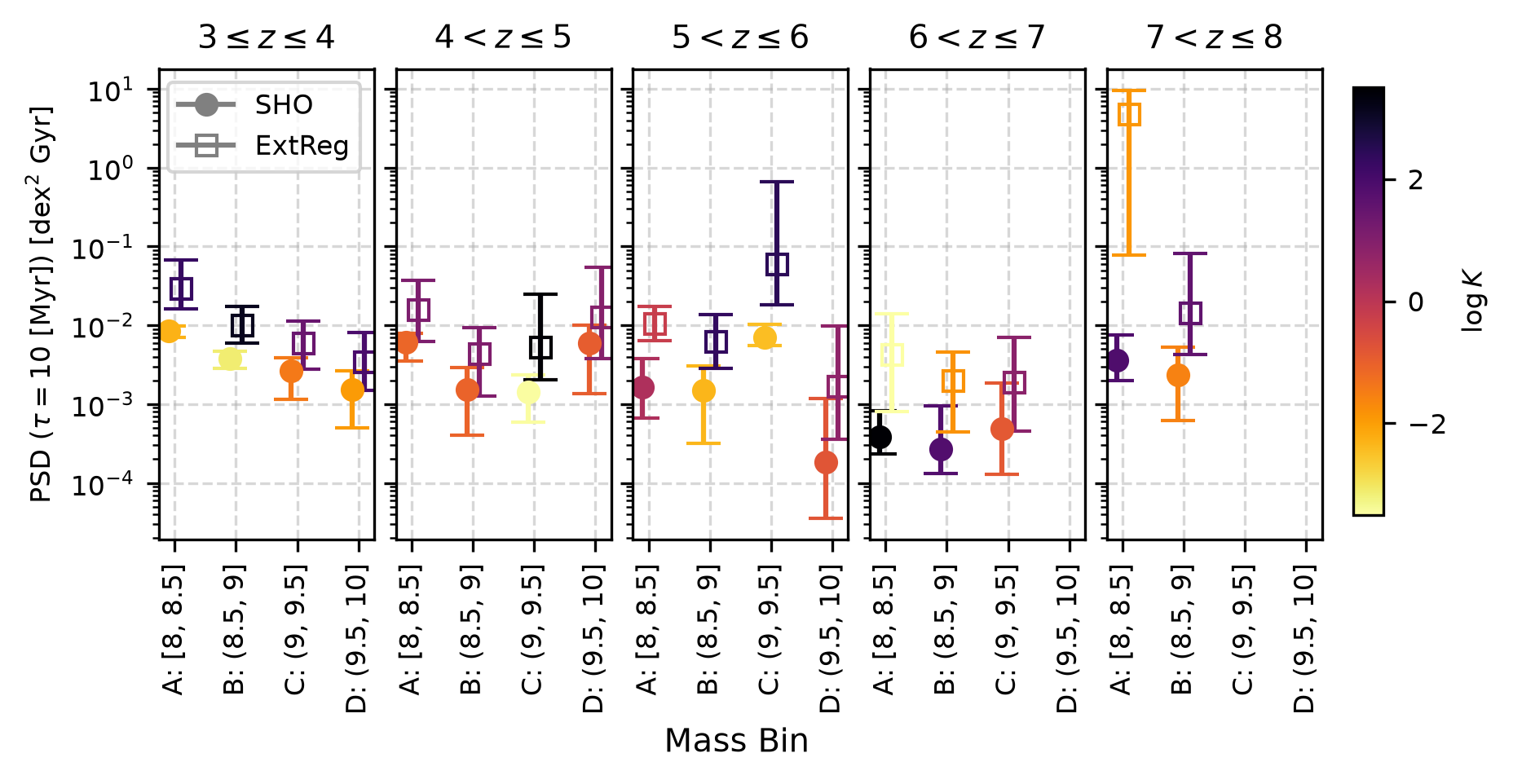}
    \caption{The Power Spectral Density at 10 Myrs inferred for each stellar mass (x axis) and redshift (panels from left to right) bin and kernel (\texttt{SHO}-circles; \texttt{ExtReg}-squares). The colour bar represents the Bayes factor $\log K$ between the two models (e.g. the ExtReg data points are coloured based on $\log K_{\rm ExtReg-SHO} = \log Z_{\rm ExtReg} - \log Z_{\rm SHO}$ and the SHO data points are coloured based on $-\log K_{\rm ExtReg - SHO}$). The error bars are the 1 sigma uncertainty inferred from the posterior samples. In the \texttt{SHO} model, the inferred 10~Myr power decreases with increasing stellar mass most clearly at $z=3$--4, indicating stronger short-timescale star-formation variability in lower-mass galaxies. This trend is weaker or less evident in the \texttt{ExtReg} model and at higher redshift.}
    \label{fig:psd10myrs}
\end{figure*}

We find some preference for one model over another as can be seen in \cref{fig:model_comparison} which shows $\Delta \log \mathcal{Z} = \log \mathcal{Z}_{\rm \texttt{ExtReg}} - \log \mathcal{Z}_{\rm \texttt{SHO}}$ as a function of stellar mass and redshift bin. The log Bayes ratio between the two models ranges in values from $\approx 3$ to $\approx -3$ nats with a weak trend in redshift in the low stellar mass bin. The variability of star formation in galaxies at $z = 4-6$ seems to be better described by the \texttt{ExtReg} model and $z=6-8$ by the \texttt{SHO} model in the $\log M_* / M_\odot = 8-8.5$ bin. However, the trend is not very consistent with stellar mass, and the population sizes vary across the data bins (see \cref{fig:galaxy-numbers}).

The strongest preference for the \texttt{ExtReg} model occurs in the best-sampled bin
($z=3$--$4$, $\log M_*/M_\odot = 8$--$8.5$; 3957 galaxies), with
$\Delta\log\mathcal{Z} = 2.28$ (betting odds of 9.8:1). The strongest preference for the
\texttt{SHO} model occurs at $z=6$--$7$ in the same mass bin (797 galaxies), with
$\Delta\log\mathcal{Z} = -3.50$ (33.1:1). We caution against interpreting this as
physical evolution in the PSD, for two reasons.

First, the number of galaxies per bin falls steeply with redshift (\cref{fig:galaxy-numbers}) so the high redshift scatters have larger uncertainties. The less informative data cannot support the additional component in the \texttt{ExtReg} model and the Occam penalty in the evidence means that there is a preference for the simpler \texttt{SHO} model.

Secondly, the high-redshift, low-mass bins are subject to a selection effect that preferentially includes galaxies observed during a burst. Galaxies undergoing a burst are bright in the rest-frame UV and have strong emission lines which boosts their detectability where as galaxies that are experiencing a lull in star formation fall below the detection limit of the survey. Furthermore, young stellar populations can outshine older populations biasing the inferred stellar mass low. Together these effects should enhance $\sigma_{\rm MS}(10~{\rm Myr})/\sigma_{\rm MS}(100~{\rm Myr})$ at high redshift and low mass which is exactly what we observe. The redshift $7 -8$ and $\log M_* / M_\odot = 8 - 8.5$ bin, which contains 234 galaxies and has a Bayes ratio of 1.88 in favour of the \texttt{SHO} model, has a much more extreme ratio of $\sigma_{\rm MS}(10~{\rm Myr})/\sigma_{\rm MS}(100~{\rm Myr}) = 2.91$ than the other bins, which have $\sigma_{\rm MS}(10~{\rm Myr})/\sigma_{\rm MS}(100~{\rm Myr}) = 1.49 - 2.20$, indicating a much more bursty population.

Together these selection effects might bias the samples towards samples of galaxies that are particularly bursty and bright at high redshift whose SFRs can be better modelled by the \texttt{SHO} PSD but do not necessarily represent the main sequence population.

\cref{fig:psd10myrs} shows the inferred PSD power on 10~Myr timescales as a function of stellar mass and redshift. In the \texttt{SHO} model, the $z=3$--4 bin shows a clear decrease in short-timescale power with increasing stellar mass, implying that lower-mass galaxies exhibit more rapid, bursty star-formation variability than higher-mass systems. This is consistent with \cite{Munoz2026psd}, who find that star-formation burstiness increases towards lower halo masses using H$\alpha$/UV ratios at $z \sim  4 - 6$, and who predict a correspondingly broader SFMS at low stellar mass. The trend is less evident in the \texttt{ExtReg} model and in the higher-redshift bins, where the current uncertainties and sample sizes limit our ability to identify a robust stellar-mass dependence. At fixed stellar mass, we find no clear monotonic evolution of the 10~Myr power with redshift. Similarly, the ratio of the inferred power on 10 and 50~Myr timescales shows no strong systematic dependence on either stellar mass or redshift, suggesting that the shape of the PSD over the timescales directly probed by the data is broadly similar across the sample.

\subsection{Main Sequence Star Formation}

\Cref{tab:extreg_params} and \cref{tab:sho_params} show the mean and one sigma standard deviations on the \texttt{ExtReg} and \texttt{SHO} PSD parameter posteriors, and \cref{fig:parameter-kl-divergence} shows the KL divergence between the 1D parameter posterior and priors for each redshift, stellar mass bin and PSD parameter. The KL divergence is highest for $\log S_0$, $\log \tau_0$, $\log \sigma_{\rm \texttt{Dyn}}$ and $\log \tau_{\rm \texttt{Dyn}}$ and close to 0 consistently for $\log \tau_{\rm \texttt{eq}}$ and $\log \tau_{\rm \texttt{in}}$. This is reflected in the posteriors themselves in \cref{fig:parameter_constraints_all_mass_redshift_bins}.

The lack of constraining power on the characteristic timescales for gas in flow and equilibriation, $\tau_{\rm \texttt{in}}$ and $\tau_{\rm \texttt{eq}}$ is because these processes happen on longer time scales than the longest timescale probed by our observations 100 Myr. The variance of the regulator component of the \texttt{ExtReg} model $\sigma_{\rm \texttt{Reg}}$ is somewhat constrained around $1$, however there remains a non-zero probability that the variance is $<1$. In contrast, the dynamical component of the \texttt{ExtReg} model is well constrained with an amplitude around $\log \sigma_{\rm \texttt{Dyn}} \approx 0$ and a timescale $\tau_{\rm \texttt{Dyn}} \sim 2 - 20$ Myr.

The \texttt{SHO} amplitude is tightly constrained around $S_0 = 0.01$ and the timescale around $\tau_0 \sim 5 - 40 $ Myr. The similarity in the constraints on $\tau_{\rm \texttt{Dyn}}$ and $\tau_0$ suggest that the \texttt{SHO} and dynamical components of the \texttt{ExtReg} model are modelling the same stochastic processes. Indeed, the lack of constraint in $\tau_{\rm \texttt{eq}}$ and $\tau_{\rm \texttt{in}}$ indicates that the regulator component is only contributing to the magnitude of the PSD at the time scales probed by the data but that the scatter in star formation on 10-100 Myr timescales is predominantly being driven by dynamical processes in the galaxy like the formation of giant molecular clouds, spiral arms or bar instabilities. 

\begin{figure*}
    \centering
    \includegraphics[width=\linewidth]{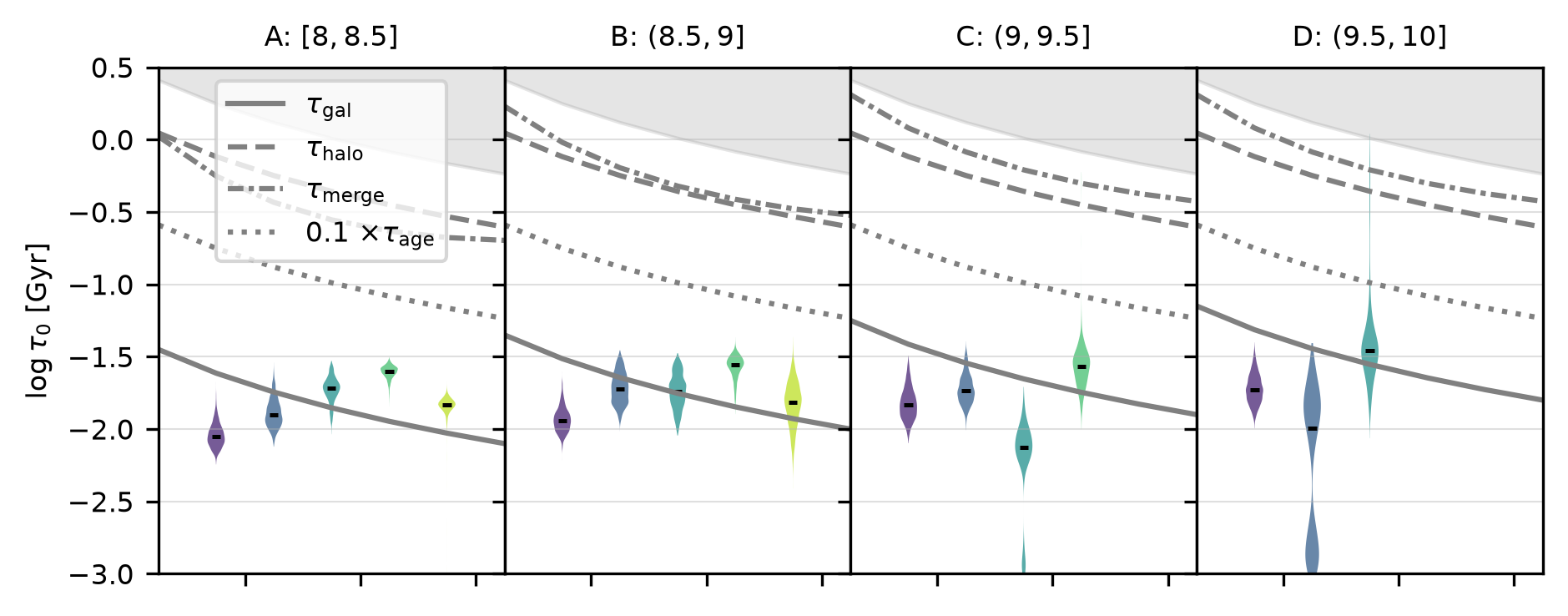}
    \includegraphics[width=\linewidth]{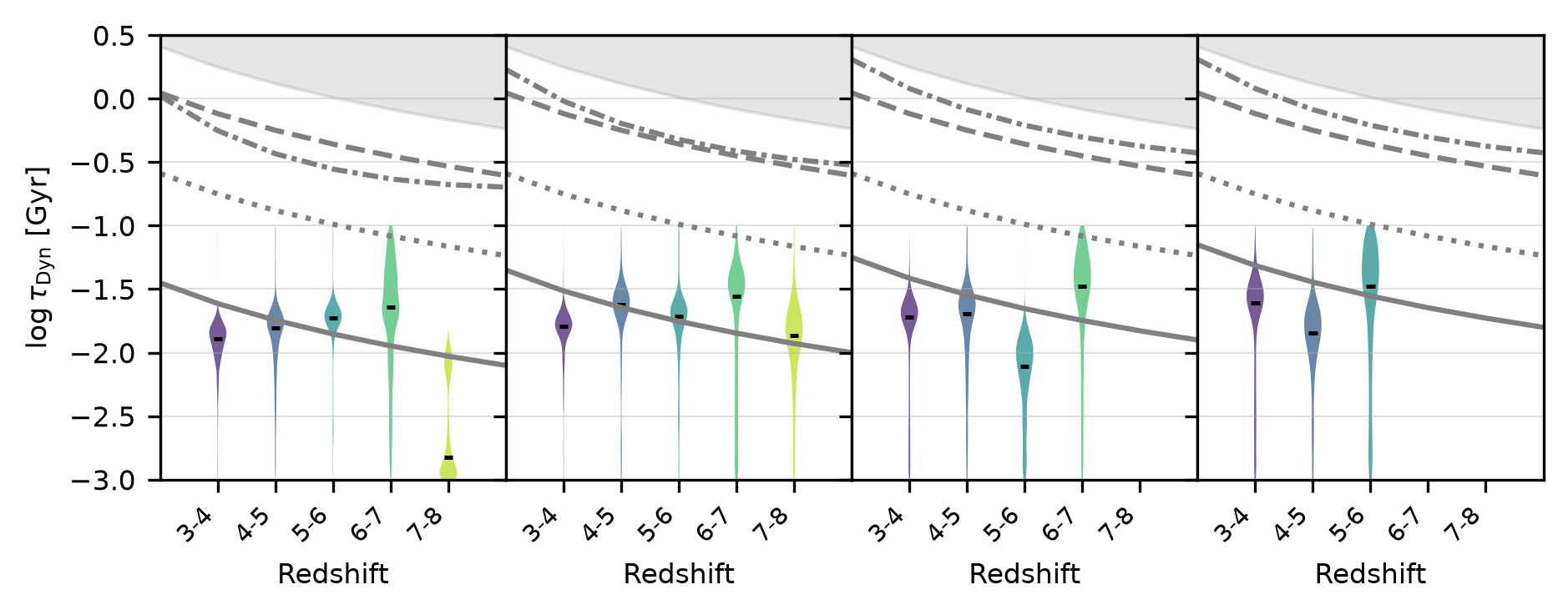}
    \caption{The constraints on $\tau_0$ and $\tau_{\rm Dyn}$ as a function of stellar mass and redshift bin and compared to different characteristic timescales. The shaded grey region shows the age of the Universe $\tau_{\rm age}$, the dotted line shows $0.1 \tau_{\rm age}$, the solid grey line show the dynamical timescale for the galaxy, the dashed line shows the halo dynamical timescale and the dash dotted line shows the merger timescale inferred from \protect\cite{Puskas2025}.}
    \label{fig:timescales}
\end{figure*}

\cref{fig:timescales} shows the constraints on $\tau_0$ and $\tau_{\rm dyn}$ as a function of redshift and stellar mass compared to several theoretically motivated characteristic timescales. The halo dynamical timescale 
\begin{equation}
    \tau_{\rm halo} = 1.4~{\rm Gyr} \bigg( \frac{1 + z}{3}\bigg)^{-1.5},
\end{equation}
is shown as a dashed line, following the standard virial scaling $\tau_{\rm dyn}\propto \rho_{\rm vir}^{-1/2}\propto H(z)^{-1}$, which at high redshift gives approximately $\tau_{\rm dyn}\propto (1+z)^{-3/2}$ \citep[e.g.][]{Dekel2009}. We also include the galactic dynamical timescale,
\begin{equation}
    \tau_{\rm gal} = 100~{\rm Myr} \bigg(\frac{M_*}{10^{10}M_\odot}\bigg)^{0.2} \bigg(\frac{1 + z}{3}\bigg)^{-1.5},
\end{equation}
as a solid grey line. This scaling captures the expectation that the dynamical times of high-redshift star-forming galaxies are short, of order tens to hundreds of Myr, and decrease with redshift as galaxies become denser \citep[e.g.][]{Dekel2009,Tacchella2016}. Such timescales are also comparable to the lifecycle of star-forming gas structures and molecular clouds, which are expected to regulate short-timescale SFR fluctuations through cloud formation, collapse, disruption, and stellar feedback \citep[e.g.][]{Krumholz2012,Kruijssen2019,Tacchella2020ExtReg}. Timescales longer than the age of the Universe are shaded out, and $0.1\times\tau_{\rm age}$ is shown as a dotted line for reference.

Finally, the merger timescale $\tau_{\rm merge}$, shown as a dot dashed line, is given by inverting the merger rate relationship
\begin{equation}
    R_M (z, M_*) = R_0 \times ( 1 + z)^{m} \exp(\tau (1+z))
\end{equation}
from \cite{Puskas2025} where $R_0$, $m$ and $\tau$ are stellar mass dependent parameters whose values are given in Table 4 of \cite{Puskas2025}.

The inferred $\tau_0$ and $\tau_{\rm Dyn}$ are generally closest to $\tau_{\rm gal}$, and are substantially shorter than the halo dynamical and merger timescales. This supports the interpretation that the 10--100~Myr scatter of the high-redshift star-forming main sequence is primarily tracing internal, dynamical star-formation variability within galaxies, rather than variability driven directly by cosmological gas accretion or galaxy mergers.

For $\tau_{\rm dyn}$ the posteriors are quite broad with long tails and while the mean of the posteriors increases with redshift our data does not rule out the $(1 + z)^{-1.5}$ trend in $\tau_{\rm gal}$. However, for $\tau_0$ the posteriors are narrower and in the lowest mass bin $\tau_0$ increases with redshift up until $z=7-8$ at which point it begins to decrease again. While this appears to be in contrast with $\tau_{\rm gal}$, we note that $\tau_{\rm gal}$ is only an order of magnitude scaling and that $\tau_0$ stays within 5 to 40 Myr of its value. Furthermore, a characteristic timescale of 10-30 Myr is consistent with the stellar feedback cycle which is set by stellar evolutionary timescales that are not expected to scale strongly with redshift. Indeed, in the high mass bins there is no obvious evolution in $\tau_0$.

\cite{Munoz2026psd} fitted a damped random walk power spectral density model to observations of UV luminosity functions, galaxy clustering and the distribution of H$\alpha$/UV ratios in UV magnitude bins at $z\sim 4-6$. They constrained the characteristic timescale for star formation in their observed galaxies to be $\approx 20$ Myr consistent with our results (although we note that there is some overlap in the data used). Furthermore, they found no evolution in the characteristic timescale for star formation with redshift consistent with what we find in our high mass bins for $\tau_0$.

The data clearly prefers a power spectral density with a characteristic timescale of $\approx 10 - 30$ Myr, and we do not see any clear trend in the model parameters as a function of redshift and stellar mass bin. This suggests that there is little evolution in the processes driving star formation on timescales between 10 and 100 Myr in our galaxy samples across $z = 3 -8$ and $\log M_* / M_\odot = 8 - 10$.

\begin{table*}
    \centering
    \begin{tabular}{llccccc}
    \hline\hline
    Stellar Mass & Redshift & $\log_{10} \sigma_{\rm \texttt{Reg}}$ & $\log_{10} \tau_{\rm \texttt{eq}}$ [Gyr]& $\log_{10} \tau_{\rm \texttt{in}}$ [Gyr] & $\log_{10} \sigma_{\rm \texttt{Dyn}}$ & $\log_{10} \tau_{\rm \texttt{Dyn}}$ [Gyr]\\
    \hline
    \hline
    $8 \leq \log M_* / M_\odot \leq 8.5$ & $3 \leq z \leq 4$ & $-0.956 \pm 0.593$ & $-1.166 \pm 1.220$ & $-1.082 \pm 0.608$ & $0.093 \pm 0.262$ & $-1.945 \pm 0.217$ \\
     & $4 < z \leq 5$ & $-0.626 \pm 0.677$ & $-1.373 \pm 1.263$ & $-1.257 \pm 0.588$ & $-0.173 \pm 0.598$ & $-1.900 \pm 0.320$ \\
     & $5 < z \leq 6$ & $-1.037 \pm 0.677$ & $-1.353 \pm 1.221$ & $-1.299 \pm 0.545$ & $-0.075 \pm 0.463$ & $-1.778 \pm 0.276$ \\
     & $6 < z \leq 7$ & $-0.218 \pm 0.541$ & $-1.950 \pm 0.997$ & $-1.563 \pm 0.436$ & $-0.449 \pm 0.576$ & $-1.724 \pm 0.453$ \\
     & $7 < z \leq 8$ & $-0.831 \pm 0.518$ & $-1.069 \pm 1.306$ & $-0.922 \pm 0.626$ & $0.415 \pm 0.172$ & $-2.610 \pm 0.391$ \\
    \hline
    $8.5 < \log M_* / M_\odot \leq 9$ & $3 \leq z \leq 4$ & $-0.921 \pm 0.612$ & $-1.165 \pm 1.245$ & $-1.130 \pm 0.592$ & $-0.097 \pm 0.374$ & $-1.853 \pm 0.241$ \\
     & $4 < z \leq 5$ & $-0.763 \pm 0.663$ & $-1.300 \pm 1.208$ & $-1.179 \pm 0.573$ & $-0.198 \pm 0.479$ & $-1.717 \pm 0.355$ \\
     & $5 < z \leq 6$ & $-0.707 \pm 0.679$ & $-1.558 \pm 1.171$ & $-1.263 \pm 0.579$ & $-0.270 \pm 0.542$ & $-1.801 \pm 0.342$ \\
     & $6 < z \leq 7$ & $-0.394 \pm 0.621$ & $-1.477 \pm 1.083$ & $-1.334 \pm 0.571$ & $-0.557 \pm 0.721$ & $-1.811 \pm 0.530$ \\
     & $7 < z \leq 8$ & $-0.647 \pm 0.659$ & $-1.122 \pm 1.240$ & $-1.058 \pm 0.615$ & $-0.038 \pm 0.257$ & $-1.938 \pm 0.364$ \\
    \hline
    $9 < \log M_* / M_\odot \leq 9.5$ & $3 \leq z \leq 4$ & $-0.786 \pm 0.649$ & $-1.309 \pm 1.202$ & $-1.158 \pm 0.598$ & $-0.284 \pm 0.567$ & $-1.868 \pm 0.395$ \\
     & $4 < z \leq 5$ & $-0.601 \pm 0.639$ & $-1.179 \pm 1.185$ & $-1.147 \pm 0.591$ & $-0.175 \pm 0.445$ & $-1.815 \pm 0.389$ \\
     & $5 < z \leq 6$ & $-0.902 \pm 0.617$ & $-1.124 \pm 1.190$ & $-1.097 \pm 0.583$ & $0.089 \pm 0.202$ & $-2.195 \pm 0.324$ \\
     & $6 < z \leq 7$ & $-0.448 \pm 0.708$ & $-1.238 \pm 1.161$ & $-1.117 \pm 0.576$ & $-0.286 \pm 0.578$ & $-1.616 \pm 0.471$ \\
    \hline
    $9.5 < \log M_* / M_\odot \leq 10$ & $3 \leq z \leq 4$ & $-0.663 \pm 0.656$ & $-1.515 \pm 1.104$ & $-1.237 \pm 0.556$ & $-0.286 \pm 0.544$ & $-1.776 \pm 0.448$ \\
     & $4 < z \leq 5$ & $-0.719 \pm 0.642$ & $-1.214 \pm 1.236$ & $-1.126 \pm 0.603$ & $-0.142 \pm 0.519$ & $-1.914 \pm 0.358$ \\
     & $5 < z \leq 6$ & $-0.415 \pm 0.728$ & $-1.162 \pm 1.167$ & $-1.045 \pm 0.561$ & $-0.296 \pm 0.597$ & $-1.659 \pm 0.537$ \\
    \hline
    \end{tabular}
    \caption{Posterior means and standard deviations in log space for the \texttt{ExtReg} parameters for different stellar mass bins and redshifts.}
    \label{tab:extreg_params}
\end{table*}

\begin{table*}
    \centering
    \begin{tabular}{llccc}
    \hline\hline
    Stellar Mass & Redshift & $\log_{10} S_0$ & $\log_{10} \tau_0 [Gyr]$ & $\log_{10} Q$ \\
    \hline
    $8 \leq \log M_* / M_\odot \leq 8.5$ & $3 \leq z \leq 4$ & $-1.663 \pm 0.046$ & $-2.043 \pm 0.081$ & $-0.189 \pm 0.088$ \\
     & $4 < z \leq 5$ & $-1.572 \pm 0.065$ & $-1.890 \pm 0.097$ & $-0.122 \pm 0.134$ \\
     & $5 < z \leq 6$ & $-1.761 \pm 0.115$ & $-1.730 \pm 0.089$ & $0.156 \pm 0.202$ \\
     & $6 < z \leq 7$ & $-1.839 \pm 0.106$ & $-1.608 \pm 0.046$ & $0.417 \pm 0.175$ \\
     & $7 < z \leq 8$ & $-2.121 \pm 0.122$ & $-1.840 \pm 0.074$ & $0.502 \pm 0.138$ \\
    \hline
    $8.5 < \log M_* / M_\odot \leq 9$ & $3 \leq z \leq 4$ & $-1.804 \pm 0.052$ & $-1.937 \pm 0.082$ & $-0.176 \pm 0.099$ \\
     & $4 < z \leq 5$ & $-1.688 \pm 0.147$ & $-1.717 \pm 0.107$ & $-0.006 \pm 0.236$ \\
     & $5 < z \leq 6$ & $-1.799 \pm 0.144$ & $-1.740 \pm 0.121$ & $0.027 \pm 0.249$ \\
     & $6 < z \leq 7$ & $-1.770 \pm 0.148$ & $-1.574 \pm 0.081$ & $0.358 \pm 0.239$ \\
     & $7 < z \leq 8$ & $-1.906 \pm 0.155$ & $-1.828 \pm 0.153$ & $0.032 \pm 0.237$ \\
    \hline
    $9 < \log M_* / M_\odot \leq 9.5$ & $3 \leq z \leq 4$ & $-1.761 \pm 0.094$ & $-1.825 \pm 0.109$ & $-0.100 \pm 0.166$ \\
     & $4 < z \leq 5$ & $-1.670 \pm 0.119$ & $-1.723 \pm 0.103$ & $-0.104 \pm 0.193$ \\
     & $5 < z \leq 6$ & $-1.876 \pm 0.200$ & $-2.200 \pm 0.293$ & $-0.119 \pm 0.130$ \\
     & $6 < z \leq 7$ & $-1.555 \pm 0.239$ & $-1.570 \pm 0.157$ & $0.157 \pm 0.281$ \\
    \hline
    $9.5 < \log M_* / M_\odot \leq 10$ & $3 \leq z \leq 4$ & $-1.685 \pm 0.130$ & $-1.722 \pm 0.102$ & $-0.036 \pm 0.215$ \\
     & $4 < z \leq 5$ & $-1.890 \pm 0.158$ & $-2.228 \pm 0.505$ & $0.183 \pm 0.304$ \\
     & $5 < z \leq 6$ & $-1.356 \pm 0.442$ & $-1.404 \pm 0.322$ & $0.172 \pm 0.290$ \\
    \hline
    \end{tabular}
    \caption{Posterior means and standard deviations for the \texttt{SHO} model parameters.}
    \label{tab:sho_params}
\end{table*}

\section{Conclusions}
\label{sec:conclusions}

We have conducted a comprehensive study of star-formation variability on the star-forming main sequence during cosmic noon and the epoch of reionization ($z = 3{-}8$) using power spectral density (PSD) models. We constrained two PSD models, the Simple Harmonic Oscillator \citep[\texttt{SHO};][]{ForemanMackey2017SHO} and the extended regulator \citep[\texttt{ExtReg};][]{Tacchella2020ExtReg}, using measurements of the intrinsic scatter of star formation around the main sequence from \cite{Simmonds2025burstingattheseams} over timescales of 10 to 100 Myr. We used Bayesian inference to explore the parameter space and trained neural network emulators to approximate the relationship between PSD parameters and scatter.

We found only weak evidence for a preference between the \texttt{SHO} and \texttt{ExtReg} models, with Bayes factors generally too small to draw strong conclusions. In the lowest stellar-mass bin, $\log M_* / M_\odot = 8$--8.5, the data tentatively favour the two-component \texttt{ExtReg} model at $z=3$--6 and the single-component \texttt{SHO} model at higher redshift, although this trend may be affected by sample size and selection effects. In addition, the \texttt{SHO} model shows a decrease in 10~Myr PSD power with increasing stellar mass, most clearly at $z=3$--4, suggesting that lower-mass galaxies exhibit stronger short-timescale star-formation variability than higher-mass systems. This is consistent with the physical expectation that shallower potential wells and less stable gas reservoirs make low-mass galaxies more susceptible to feedback-driven fluctuations.

We are unable to constrain the timescales of gas cycling and infall, with the regulator component of the \texttt{ExtReg} model contributing primarily to the overall amplitude of the PSD inferred from the observations. In contrast, the observed intrinsic scatters between 10 and 100~Myr constrain both the dynamical component of the \texttt{ExtReg} model and the \texttt{SHO} model to characteristic timescales of $\simeq 10$--30~Myr. Our inferred characteristic timescales of $\simeq 10$--30~Myr are consistent with the timescales on which massive stars regulate their surrounding ISM through winds, radiation pressure, photoionization, and supernova feedback \cite[e.g.,][]{Chevance2020GMCs, Munoz2026psd}. In this picture, short-timescale SFMS scatter reflects the response of the gas reservoir to recent star formation: gas collapses and forms stars, feedback suppresses or disrupts subsequent star formation, and the system then recovers as gas re-accumulates or re-cools. Such feedback-regulated cycling provides a natural physical origin for enhanced variability on $\sim 10$~Myr timescales, particularly in low-mass galaxies where shallow potential wells make the ISM more susceptible to feedback-driven perturbations. This interpretation is consistent with \citet{Simmonds2025burstingattheseams}, who found enhanced SFMS scatter on short SFR-averaging timescales at high redshift, and with recent independent evidence for broadly similar levels of star-formation burstiness across $z\sim3$--7 \citep[see also][]{Mitsuhashi2026}.

Observations of the main sequence scatter on longer timescales are needed to constrain $\tau_{\rm eq}$ and $\tau_{\rm \texttt{in}}$ or some alternative tracer of star formation on longer timescales. While this work has focused on constraining the parameters of the star-formation PSD from a population level summary statistic $\sigma_{\rm MS}$, additional information about star formation will be available in the individual galaxy SFHs and spectra. One possible approach to constraining the power spectral density of star formation on longer timescales would be to look at the distribution of spectral line fluxes and equivalent widths in a population of galaxies. \cite{Iyer2024ExtReg} showed that distributions of H$\alpha$, H$\delta$, Ca-H and K absorption, $D_n(4000)$, UV indicators and broadband colours in a population of galaxies carry complementary signatures of the stochasticity in star formation and can be used to constrain the parameters of the \texttt{ExtReg} model. Another approach would be to look at the impact of the PSD parameters on the distribution of ratios like ${\rm SFR}_{t_{\rm avg} = 10~{\rm Myr}} / {\rm SFR}_{t_{\rm avg} = 100~{\rm Myr}}$ for individual galaxies or equivalently the ratio of H$\alpha$ to UV as is done in \cite{Munoz2026psd}. Mapping PSD parameters to a distribution of galaxy statistics (equivalent widths or time averaged SFRs) would however require some form of implicit likelihood inference and neural density estimation. This is left for future work.

\section*{Acknowledgements}

HTJB acknowledges support from the Kavli Institute for Cosmology Cambridge and the Kavli Foundation.

\section*{Data Availability}

The code is available at \url{https://github.com/htjb/SFV-SFMS} and the emulators and inference chains are available at \url{https://doi.org/10.5281/zenodo.20595727}. The data is available upon reasonable request to the authors.



\bibliographystyle{mnras}
\bibliography{example} 

\appendix

\section{Halo Seed Mass and Observed Stellar Mass}

\begin{figure}
    \centering
    \includegraphics[width=\linewidth]{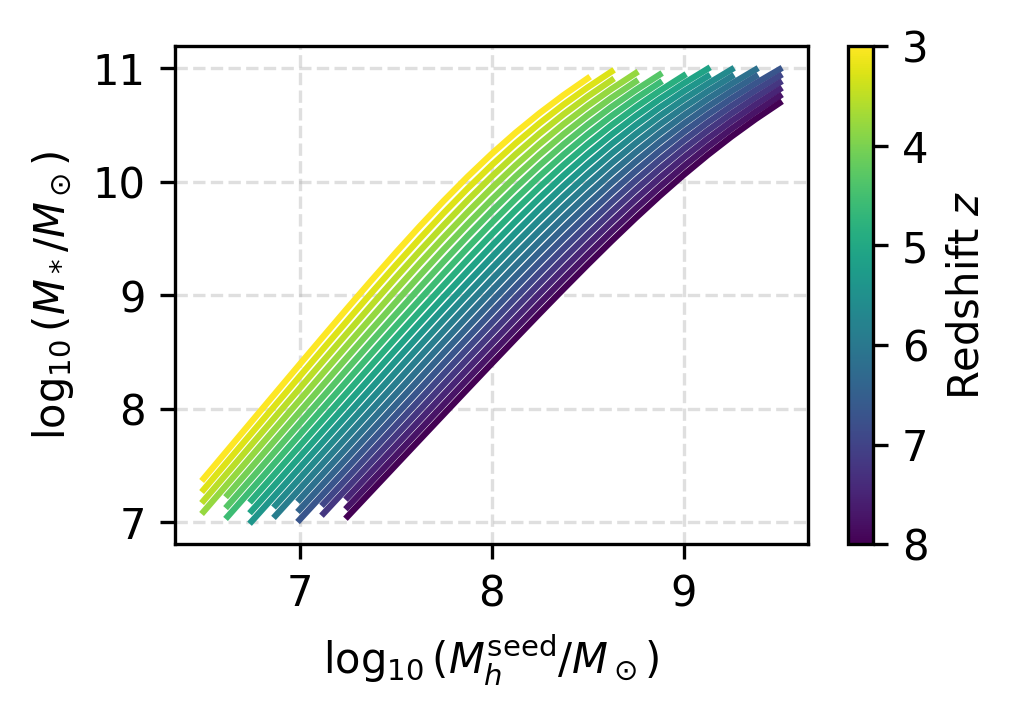}
    \caption{In order to generate catalogues of galaxies with a given stellar mass at a given redshift to train the emulators used in this work the seed halo mass of the galaxies at high redshift needs to be chosen carefully. To pick an appropriate seed halo mass we explore and interpolate the relationship, shown in the figure, between seed halo mass of a galaxy at redshift 30, the observed redshift and observed stellar mass inferred from \cref{eq:accretion-rate} and \cref{eq:star-formation-efficiency}.}
    \label{fig:seed-mass-stellar-mass}
\end{figure}

For each stellar mass and redshift bin with an observed intrinsic scatter, we train an emulator on simulations of galaxies that fall within the same bin. Under our model each galaxy is assumed to have the same mean SFH given by ${\rm SFR} (M_h, z) = f_b \epsilon_*(M_h) \dot{M_h}$ as discussed in \cref{sec:simulations}. To get the halo mass history, we integrate the halo mass accretion rate $\dot{M_h}$ from redshift 30 to the observation redshift $z_{\rm obs}$. The average stellar mass at $z_{\rm obs}$ is therefore determined by the initial halo mass at $z=30$ and we interpolate the relationship between observation redshift, seed halo mass and final stellar mass shown in \cref{fig:seed-mass-stellar-mass} to ensure our simulated galaxies fall within the appropriate observation stellar mass bin.

\section{The 1D Parameter Posteriors}

\begin{figure*}
    \centering
    \includegraphics[width=0.8\linewidth]{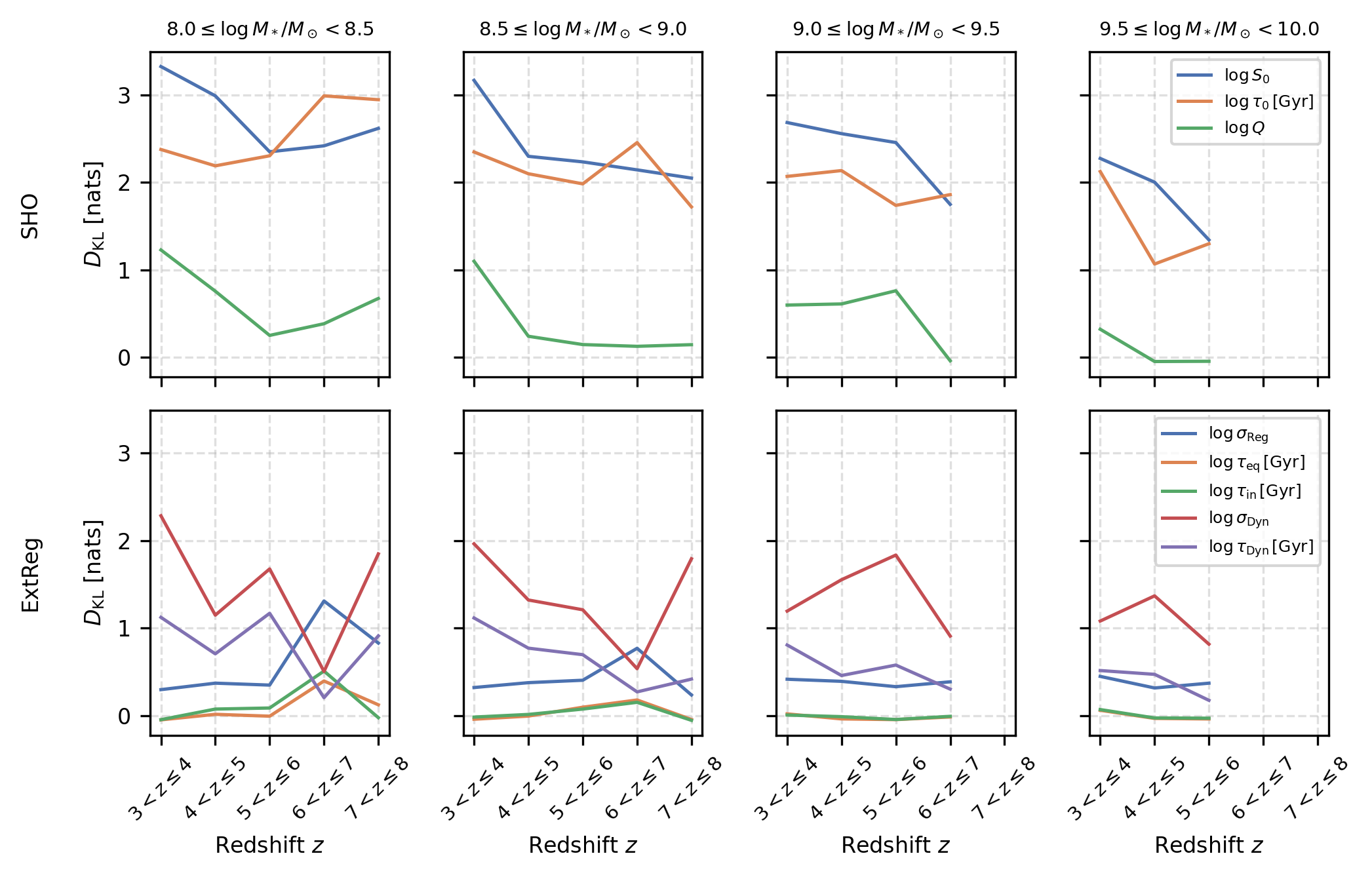}
    \caption{The Kullback-Leibler divergences between the posterior and prior distributions for each parameter, stellar mass and redshift bin and each PSD model. $\tau_{\rm \texttt{eq}}$ and $\tau_{\rm \texttt{in}}$ are notably unconstrained where as $S_0$, $\tau_0$, $\sigma_{\rm \texttt{Dyn}}$ and $\tau_{\rm \texttt{Dyn}}$ are all constrained with higher $D_{\rm KL}$. The constraints are notably stronger for the \texttt{SHO} model due to; decreased model complexity, smaller emulator errors or both.}
    \label{fig:parameter-kl-divergence}
\end{figure*}

\begin{figure*}
    \centering
    \includegraphics[width=0.48\linewidth]{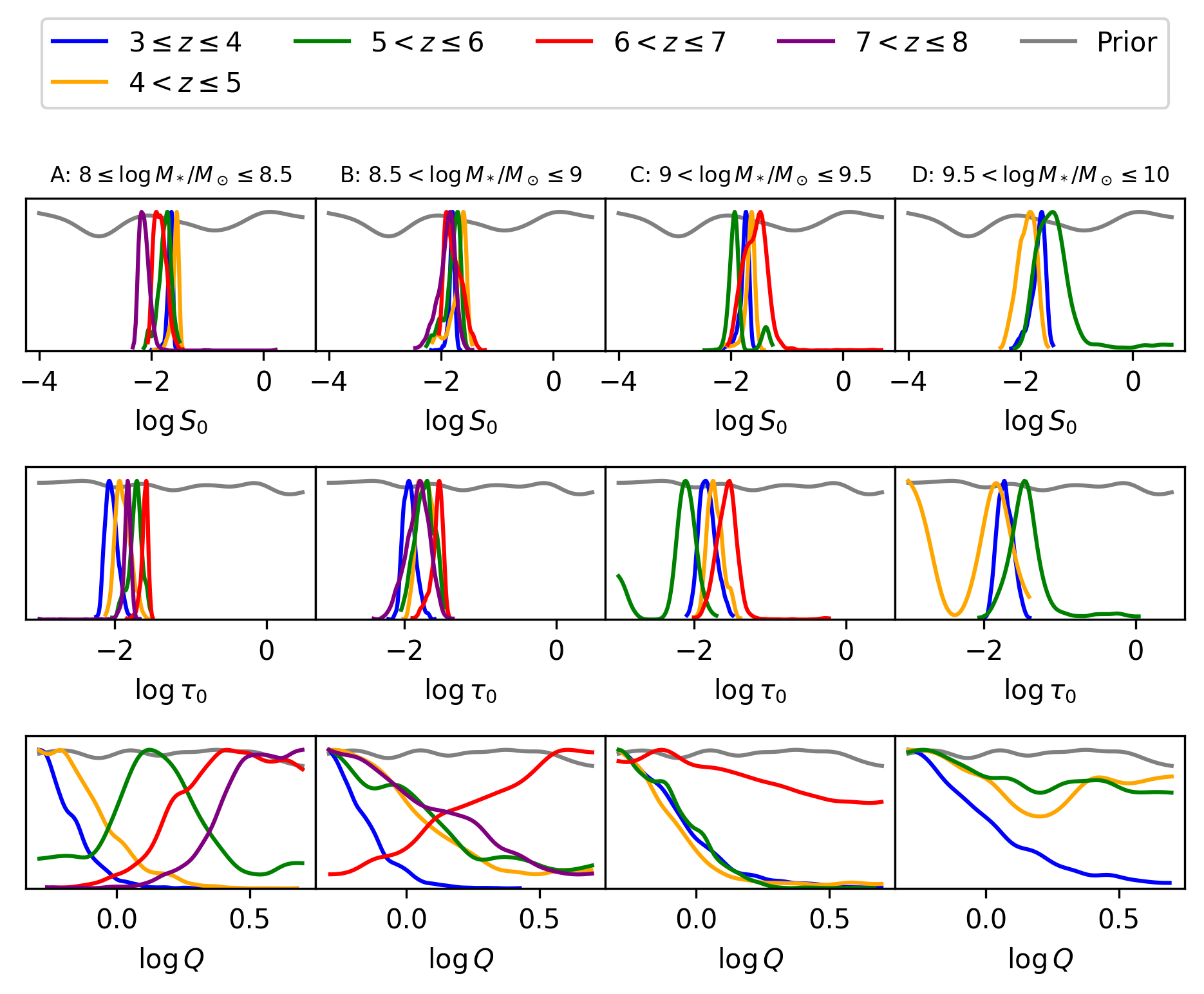}
    \includegraphics[width=0.48\linewidth]{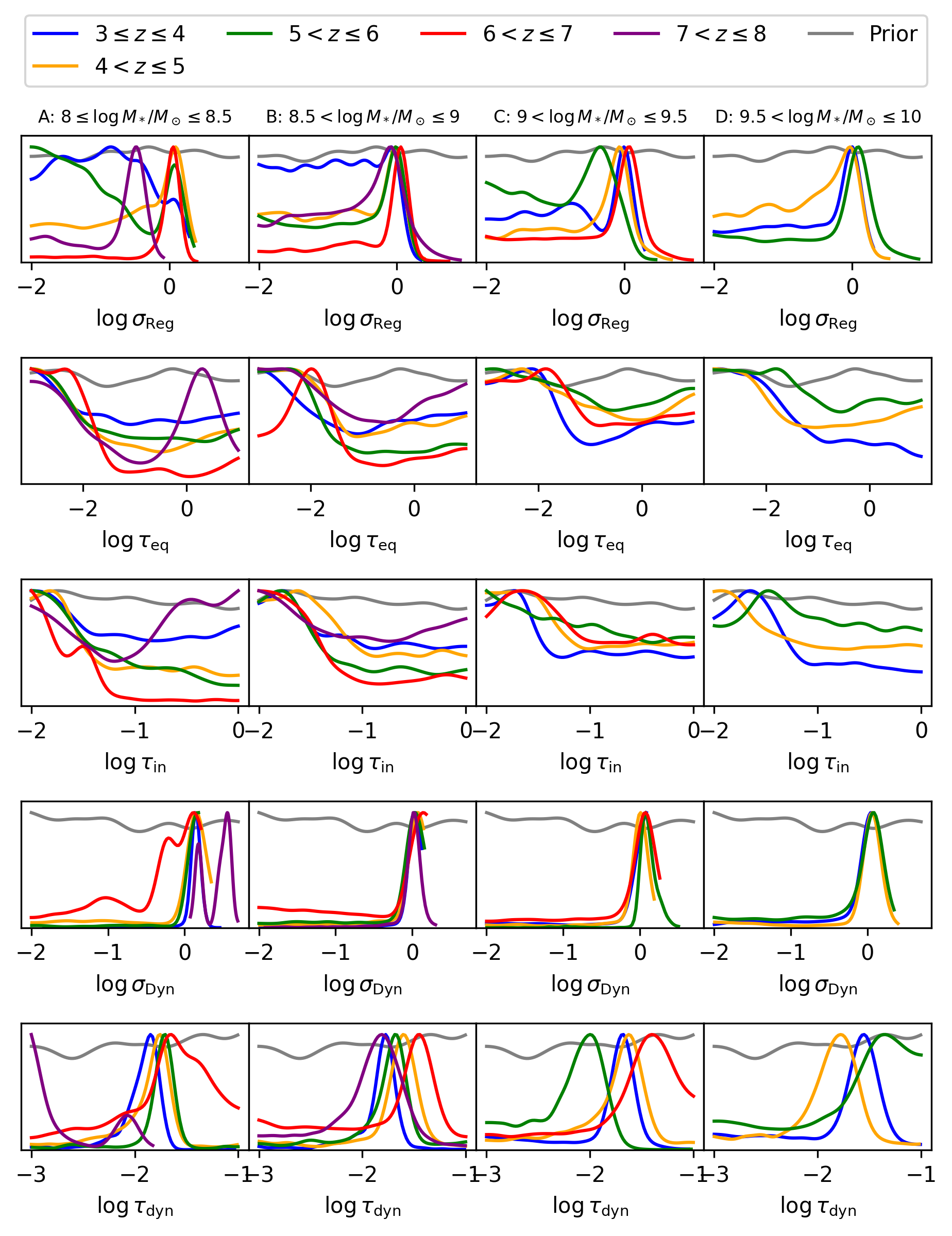}
    \caption{The 1D posteriors on each PSD parameter for the \texttt{SHO} model (left) and \texttt{ExtReg} model (right) as a function of redshift (color) and stellar mass (column) bin.}
    \label{fig:parameter_constraints_all_mass_redshift_bins}
\end{figure*}

\cref{fig:parameter-kl-divergence} shows the KL divergence between the posterior and prior for each parameter in the \texttt{SHO} and \texttt{ExtReg} models as functions of stellar mass-redshift bin. The constraints are noticeably stronger for the \texttt{SHO} model parameters and the parameters of the regulator component of the \texttt{ExtReg} model are largely unconstrained. This is further illustrated by the 1D posterior plots in \cref{fig:parameter_constraints_all_mass_redshift_bins}. See the main text for further discussion.

\bsp	
\label{lastpage}
\end{document}